\documentclass[useAMS,usenatbib]{mn2e}
\usepackage{times}
\input{psfig.sty}
\usepackage{epsfig,rotating,natbib}

\newif\ifAMStwofonts

\def\msun{M$_{\odot}$}
\def\Mo{M$_{\odot}$ }
\def\bi{\begin{itemize}}

\def\ei{\end{itemize}}
\def\be{\begin{equation}}
\def\ee{\end{equation}}
\def\bea{\begin{eqnarray}}
\def\eea{\end{eqnarray}}
\def\gcc{g cm$^{-3}$}

\begin{document}

\title[
3D Lagrangian MHD with Euler potentials
] {
MAGMA: a 3D, Lagrangian magnetohydrodynamics code for merger applications}
 
\author[Rosswog \& Price]{Stephan Rosswog$^{1}$, 
                         Daniel Price$^{2}$\\
                         ${\bf 1.}$
                         School of Engineering and Science,
                         Jacobs University Bremen (until February 2007
                         International University Bremen), 
                         Campus Ring 1, Germany\\
                         ${\bf 2.}$School of Physics, University of Exeter,
                         Stocker Road, Exeter EX4 4QL,UK.}

\date{}

\maketitle

\label{firstpage}

\begin{abstract}
We present a new, completely Lagrangian magnetohydrodynamics code that is
based on the SPH method. The equations 
of self-gravitating hydrodynamics are derived self-consistently from
a Lagrangian and account for variable smoothing length (``grad-h''-) terms
in both the hydrodynamic and the gravitational acceleration equations.
The evolution of the magnetic field is formulated in terms of so-called Euler
potentials which are advected with the fluid and thus guarantee the MHD
flux-freezing condition. This formulation is equivalent to a vector potential
approach and therefore fulfills the $\vec{\nabla}\cdot\vec{B}=0$-constraint by
construction. Extensive tests in one, two and three dimensions are presented.
The tests demonstrate the excellent conservation properties of the code 
and show the clear superiority of the Euler potentials over earlier
magnetic SPH formulations.
\end{abstract}
\begin{keywords}
methods: numerical, magnetic fields, MHD, stars: magnetic fields
\end{keywords}

\section{Introduction}

In many areas of astrophysics a transition from pure hydrodynamic
to magnetohydrodynamic simulations is underway. Magnetohydrodynamic
calculations have a long-standing tradition in the context of core-collapse
supernovae, see for example 
\citet{leblanc70,bisnovatyi_kogan76,meier76,symbalisty84}. In recent years
MHD simulations of core-collapse supernovae have seen a renaissance 
\citep[e.g.][]{akiyama03,mizuno04,liebendoerfer04,yamada04,kotake04,ardeljan05,proga05,obergaulinger06a,masada06,shibata06b,burrows07a},
mainly due to the conclusion that a small fraction of core-collapse
supernovae, those related to long gamma-ray bursts, require relativistic and
well-collimated jets and due to the difficulty to make supernovae explode via
the delayed, neutrino-driven mechanism. 
In accretion physics the magnetorotational instability
\citep[e.g.][]{balbus98} is now the widely accepted mechanism behind the
angular momentum transport that determines the accretion rate. Many of the
recent accretion simulations are performed in the framework of (sometimes 
general relativistic) magnetohydrodynamics
\citep[e.g.][]{stone01,devilliers03,sano04,mckinney05,hawley06,mckinney07}.
Several of the more recent star and planet formation calculations have also
included effects of the magnetic field
\citep[e.g.][]{hosking04,nelson05,ziegler05,machida05,fromang06,banerjee06,price07b}. 
Most recently, MHD simulations were also used in the context of compact binary
mergers \citep[e.g.][]{shibata06a,duez06,price06b,duez07}.\\  
Nearly all of the above calculations have been carried out on Eulerian
grids. Some SPH-formulations that include magnetic fields exist
\citep[e.g.][]{gingold77,phillips85,boerve01,dolag02,price05}, but 
obtaining a stable formulation has proved notoriously difficult, not least
because of the difficulty in fulfilling the
$ \; \vec{\nabla}\cdot\vec{B}=0$-constraint on Lagrangian particles (see
e.g. \cite{price07c} for a brief review). 
Nevertheless, for applications 
that involve large deformations Lagrangian schemes have definite advantages.\\
Here we present a detailed description of our Lagrangian magnetohydrodynamics
code, MAGMA (``a magnetohydrodynamics code for merger applications``). Our
developments are mainly driven by the application to mergers of 
magnetized neutron stars, but the described methods are applicable to smoothed
particle (magneto-) hydrodynamics in general. Ingredients of the code that
have been described elsewhere are briefly summarized, the 
interested reader is referred to the literature for more details. 
The focus lies on the description and the testing of the new code
elements. These are mainly improvements of the hydrodynamics part that enhance
the accuracy by a careful accounting of the so-called ``grad-h''-terms and the
inclusion of magnetic field evolution. We present a formulation of the
magnetic field evolution in terms of the so-called Euler potentials
\citep[e.g.][]{euler1769,stern94a} that are advected with the flow and thus 
guarantee the MHD flux-freezing condition. For comparison, our code also
allows to evolve the B-fields via a SPH-discretized version of the
MHD-equations  \citep{price05}.\\
The paper is organized as follows. Section \ref{sec:method} 
describes the details of 
the numerical methods and their implementation. It includes a brief summary
of code elements described previously (Sec. \ref{sec:summary_old}), a concise
summary of the hydrodynamics plus gravity as derived from a Lagrangian
accounting for the so-called ``grad-h''-terms (Sec. \ref{sec:hydro}), and 
a detailed description of the treatment of the magnetic field
(Sec. \ref{sec:magnet}). Various tests of the different code elements are
presented Sec. \ref{sec:tests}. The paper is 
summarized in Sec.~\ref{sec:summary}.

\section{Method description}
\label{sec:method}
This section describes the methods and the implementation of the 
various elements of our  new (magneto-)hydrodynamics code.
Historically, numerical MHD schemes have first been developed for grid-based
methods. Several implementations of magnetic fields into smoothed particle
hydrodynamics exist, e.g. \cite{gingold77,phillips85,boerve01}, but their
success has somewhat been hampered by several numerical difficulties,
not least of which is the difficulty in fulfilling the
$\vec{\nabla}\cdot\vec{B}=0$-constraint. 
In grid-based methods various techniques exist to enforce this constraint.
Our main reasons for choosing the smoothed particle hydrodynamics method 
are the exact conservation of mass, energy, linear and angular momentum by 
construction and its ease in treating vacuum. On top of that, a Lagrangian
scheme is a natural choice to simulate the highly variable geometry that
occurs during stellar collisions. The SPH method has been described and 
reviewed many times in the literature (e.g. 
\cite{benz90a,monaghan92,monaghan05}) and this will not be repeated here. 
In comparison with our earlier work we have made substantial changes in the 
implementation of the hydrodynamics equations that are documented and 
tested below, see Sec. \ref{sec:hydro} and \ref{sec:hydro_tests}. 
In summary, although the new and more sophisticated formulations improve 
the accuracy slightly, the changes in the results are only minor for the 
applications that we discuss  here.\\
The other major change is the inclusion of magnetic fields, which is 
described in Sec. \ref{sec:magnet} and tested in Sec. \ref{sec:mhd_tests}. 

\subsection{Summary of code ingredients described elsewhere}
\label{sec:summary_old}
The neutron star matter is modeled with the temperature-dependent 
relativistic mean-field equation of state of \cite{shen98a,shen98b}. It 
can handle temperatures from 0 to 100 MeV, electron fractions from $Y_e$= 0 
up to 0.56 and densities from about 10 to more than $10^{15}$ \gcc. No 
attempt is made to include matter constituents that are more exotic than 
neutrons and protons at high densities. For more details on this topic 
we refer to \cite{rosswog02a}.\\
The code also contains a detailed multi-flavor neutrino leakage scheme. 
An additional mesh is used to calculate the neutrino opacities that are 
needed for the neutrino emission rates at each particle position.
The free-streaming and neutrino diffusion limit are reproduced
correctly, the semi-transparent regime is treated by an interpolation between
these limiting cases. The neutrino emission rates calculated in this way are
used to account for the local cooling and the compositional changes due to 
weak interactions such as electron captures. A detailed description of the 
neutrino treatment can be found in \cite{rosswog03a}.\\
At present, the self-gravity of the fluid is treated in a Newtonian fashion.
Both the gravitational forces and the search for the
particle neighbors that are required, for example, to calculate densities 
or pressure gradients, are performed with a binary tree that is based 
on the one described in \cite{benz90b}. These tasks are the computationally 
most expensive part of the simulations and in practice they completely
dominate the CPU-time usage. The tree is parallelized and allows in its
current form the simulation of several million particles on a medium-sized (24
processor) shared-memory computer. Forces emerging from the emission of
gravitational waves are treated in a simple approximation. For more details,
we refer to \cite{rosswog00} and \cite{rosswog02a}.

\subsection{Hydrodynamics}\label{sec:hydro}
We are interested in a numerical solution of the Lagrangian, self-gravitating
Euler equations of ideal hydrodynamics:
\be
\frac{d\vec{v}}{dt}= - \frac{\nabla p}{\rho} - \nabla \Phi,
\ee
where $\vec{v}$ is the fluid velocity, $p$ the pressure, $\rho$ the mass
density and $\Phi$ the gravitational potential. The evolution equation for 
the specific internal energy, $u$, is
\be
\frac{du}{dt}= - \frac{p}{\rho} \nabla \cdot \vec{v}\label{eq:energy_eq}
\ee
and the density evolves according to
\be
\frac{d\rho}{dt}= - \rho \nabla \cdot \vec{v}.\label{eq:continuity_eq}
\ee

\subsubsection{SPH with ``grad-h''-terms}
\label{sec:hydro_gradh}
The exact conservation of energy, linear and angular momentum even 
in the discretized form of the equations is a major strength
of smoothed particle hydrodynamics. The equations of motion 
can be derived by using nothing more than a suitable Lagrangian, the first 
law of thermodynamics and a prescription on how to obtain an density 
estimate via summation. \\
The first derivation of the SPH-equations that takes terms from the 
derivatives of the smoothing lengths into account goes back to \cite{nelson94}.
More recently, \cite{springel02}  and \cite{monaghan02} 
derived the corresponding equations from a Lagrangian in two different 
ways. The equations of this more recent approach are less  cumbersome to
implement, but they need an extra iteration for each particle at each time
step to make the density and smoothing length consistent with each other. The
advantage of a derivation from a Lagrangian is -apart from its elegance- that
the  resulting equations  guarantee the conservation of the
physically conserved quantities, provided that the Lagrangian possesses 
the right symmetry properties. An in-depth analysis of various SPH-variants
can be found in \cite{price04c}. Without going into details of the derivations,
we will briefly sketch how to arrive at the SPH-equations including the 
so-called grad-h terms.\\
The Lagrangian of a perfect fluid is given by \citep{eckart60}

\be
L_{\rm hyd}= \int \rho \left(\frac{v^2}{2} - u(\rho,s) \right) dV,
\ee
where $\rho$ is the mass density, $v$ the fluid velocity, $u$ the specific
energy (``energy per mass'') and $s$ the specific entropy.
In SPH-discretization the Lagrangian reads

\be
L_{\rm SPH,h}= \sum_j m_j \left(\frac{v_j^2}{2} - u_j(\rho_j,s_j) \right),
\ee
where the indexed quantities refer to the values at the SPH particle
positions. This Lagrangian does not include gravity, the gravitational terms
will be discussed in Sec.\ref{sec:grav}. The discretized momentum equation 
is then found by applying the Euler-Lagrange equations

\be
\frac{d}{dt} \left(\frac{\partial L_{\rm SPH,h}}{\partial \vec{v}_i} \right) -
\frac{\partial L_{\rm SPH,h}}{\partial \vec{x}_i} = 0,\label{eq:EL_DGL}
\ee
where $\vec{x}_i$ and $\vec{v}_i$ refer to the position and velocity of
particle $i$. The first term in Eq.~(\ref{eq:EL_DGL}) provides the change of
the particle momentum, $m_i \dot{\vec{v}}_i$, the second term in the Lagrangian
acts like a potential. The second term in Eq.~(\ref{eq:EL_DGL}) becomes

\be
\frac{\partial L_{\rm SPH,h}}{\partial \vec{x}_i}=- \sum_j m_j \frac{\partial
  u_j(\rho_j,s_j)}{\partial \vec{x}_i}= - \sum_j m_j \left.\frac{\partial u_j}{\partial \rho_j}\right\vert_s 
\frac{\partial \rho_j}{\partial \vec{x}_i}.
\label{eq:dL_dx}
\ee
The derivative with respect to density can be expressed via the first law of
thermodynamics, which reads for the adiabatic case 

\be
du=  \frac{P}{\rho^2} d\rho,
\label{eq:first_LoT}
\ee
where $P$ is the gas pressure. Therefore,

\be
\left.\frac{\partial u_j}{\partial \rho_j}\right\vert_s= \frac{P_j}{\rho_j^2}
\label{eq:du_drho}
\ee
and the momentum equation becomes 

\be
m_i \frac{d\vec{v}_i}{dt}= - \sum_j m_j \frac{P_j}{\rho_j^2} \frac{\partial \rho_j}{\partial \vec{x}_i}.\label{eq:mi_dot_vi}
\ee
Eq.~(\ref{eq:first_LoT}) also provides us with the evolution equation for the
specific energy

\be
\frac{du_i}{dt}=  \frac{P_i}{\rho_i^2} \frac{d\rho_i}{dt}.\label{eq:du_dt}
\ee
For the explicit SPH equations we need to specify a prescription for 
the density and to calculate its derivatives 
$\frac{\partial \rho_j}{\partial \vec{x}_i}$ and $\frac{d\rho_i}{dt}$, 
see Eqs.~(\ref{eq:dL_dx}) and (\ref{eq:du_dt}). For the density we use

\be
\rho_i= \sum_j m_j W(r_{ij},h_i).\label{eq:rho}
\ee
Here, $W$ is the SPH smoothing kernel, $r_{ij}= |\vec{r}_{ij}|= 
|\vec{r}_i-\vec{r}_j|$ and $h_i$ is the smoothing length of particle $i$.
Throughout this paper we use the standard cubic spline kernel most often used
in SPH \citep{monaghan85}. 
Note that contrary to some earlier formulations of SPH, only the
smoothing length of the particle itself, $h_i$, is used rather than 
some average. To obtain adaptivity, we determine the smoothing length
evolution from the density (which for equal mass particles is equivalent to a
dependence on the particle number density). In 3D we use
\be
h_i= \eta \left(\frac{m_i}{\rho_i}\right)^{1/3},\label{eq:h}
\ee
where $\eta$ is a parameter typically in a range between 1.2 and 1.5. A
careful discussion of the choice of $\eta$ can be found in Price (2004).
The density $\rho_i$ depends on $h_i$, see Eq.~(\ref{eq:rho}), and vice versa, 
see Eq.~(\ref{eq:h}), so an iteration is required to reach
consistency. Typically we iterate until the relative change between two
iterations is smaller than $10^{-3}$.\\
By straight forward differentiation of the density sum, Eq.~(\ref{eq:rho}),
one obtains

\be
\frac{d\rho_i}{dt}= \frac{1}{\Omega_i} \sum_j m_j \vec{v}_{ij} \nabla_i
W_{ij}(h_i) \label{eq:drho_dt},
\ee
where $\vec{v}_{ij}=\vec{v}_i-\vec{v}_j$ and

\be
\frac{\partial \rho_j}{\partial \vec{x}_i}
= \frac{1}{\Omega_j} \sum_k m_k \frac{\partial W_{jk}(h_j)}{\partial 
\vec{x}_i},\label{eq:drho_dx}
\ee
where

\be
\Omega_k\equiv \left(1 - \frac{\partial h_k}{\partial \rho_k} \cdot \sum_l m_l
\frac{\partial}{\partial h_k} W_{kl}(h_k) \right).\label{eq:omega_i}
\ee
With the derivatives, Eqs.~(\ref{eq:drho_dt}) and (\ref{eq:drho_dx}), at hand
the energy equation, Eq.~(\ref{eq:du_dt}), becomes

\be
\frac{d u_{i,\rm h}}{dt}= \frac{1}{\Omega_i}\frac{P_i}{\rho_i^2}
\sum_j m_j \vec{v}_{ij} \nabla_i W_{ij}(h_i) \label{eq:energy_equation}
\ee
and the momentum equation reads

\be
\frac{d\vec{v}_{i,\rm h}}{dt}= - \sum_j m_j \left\{
\frac{P_i}{\Omega_i\rho_i^2} \nabla_i W_{ij}(h_i)
+ \frac{P_j}{\Omega_j\rho_j^2} 
\nabla_i W_{ij}(h_j) \right\}.\label{eq:momentum_equation_SPH}
\ee
We calculate the density via summation, see Eq.~(\ref{eq:rho}), which solves
the continuity equation without the need to explicitely evolve the density.\\

\noindent For reasons of reference we provide the SPH equations with 
averaged smoothing lengths, $h_{ij}= (h_i+h_j)/2$, that are still widely used
and that we have used in earlier calculations (this will be referred to
as the ``old equations'' or the $h_{ij}$-version). In this $h_{ij}$-version 
the density summation reads

\be
\left(\rho_{i}\right)_{ij} = \sum_j m_j W(|\vec{r}_i - \vec{r}_j|,h_{ij})|),
\ee
the energy equation is

\be
\left(\frac{d u_{i,\rm h}}{dt}\right)_{ij}= \frac{P_i}{\rho_i^2} \sum_j m_j \vec{v}_{ij} 
\nabla_i W_{ij}(h_{ij}) \label{eq:energy_equation_old}
\ee
and the momentum equation reads

\be
\left(\frac{{d\vec{v}}_{i,\rm h}}{dt}\right)_{ij}= - \sum_j m_j \left\{
\frac{P_i}{\rho_i^2} + \frac{P_j}{\rho_j^2} 
 \right\}\nabla_i W_{ij}(h_{ij}).\label{eq:momentum_equation_SPH_old}
\ee

\subsubsection{Self-gravity and gravitational softening}\label{sec:grav}

Most often gravitational softening is done by -physically motivated- but still
ad hoc recipes. It is, however, possible to derive the gravitational softening
terms self-consistently from a Lagrangian and to also take the effects from a 
locally varying smoothing length into account \citep{price07a},
similar to the case of the hydrodynamics equations.\\
If gravity is taken into account, a gravitational part has to be added 
to the Lagrangian, $L_{\rm SPH}= L_{\rm SPH,h} + L_{\rm SPH,g}$. This
gravitational part of the Lagrangian reads

\be
L_{\rm SPH,g}= - \sum_j m_j \Phi_j, \label{eq:Lgrav}
\ee
where $\Phi_j$ is the potential at the particle position $j$,
$\Phi(\vec{r}_j)$. The potential $\Phi$ can be written as a sum 
over particle contributions

\be
\Phi (\vec{r})= - G \sum_j m_j \phi(|\vec{r}-\vec{r}_j|,h), 
\label{eq:phi_sum_over_particles}
\ee
where $h$ is the smoothing length, $\phi$ the gravitational softening kernel,
and the potential is related to the matter density by Poisson's equation

\be
\nabla^2 \Phi= 4 \pi G \rho. \label{eq:poisson}
\ee
If we insert the sum representations of both the potential, 
Eq.~(\ref{eq:phi_sum_over_particles}), and the density, Eq.~(\ref{eq:rho})
into the Poisson equation, Eq.~(\ref{eq:poisson}), we obtain a relationship
between the gravitational softening kernel, $\phi$, and the SPH-smoothing
kernel $W$:

\be
W(|\vec{r}-\vec{r}_j|,h)= -\frac{1}{4 \pi} \frac{\partial}{\partial r} 
\left( r^2 \frac{\partial}{\partial r} \phi(|\vec{r}-\vec{r}_j|,h)\right).
\ee
Here we have used that both $\phi$ and $W$ depend only radially on the
position coordinate. \\
Applying the Euler-Lagrange equations, Eq.~(\ref{eq:EL_DGL}), 
to $L_{\rm SPH,g}$ yields the particle acceleration due to gravity 
\citep{price07a}

\bea
\frac{{d\vec{v}}_{i,\rm g}}{dt}&=&  
-G \sum_j m_j \left[ \frac{\phi'_{ij}(h_i)+\phi'_{ij}(h_j)}{2}\right]
\hat{e}_{ij} \nonumber\\
&-& \frac{G}{2} \sum_j m_j 
\left[ 
\frac{\zeta_i}{\Omega_i} \nabla_i W_{ij}(h_i)+
\frac{\zeta_j}{\Omega_j} \nabla_i W_{ij}(h_j)\label{eq:dv_dt_grav}
\right],  
\eea
where $\phi'_{ij}= \partial \phi/\partial |\vec{r}_i - \vec{r}_j|$ and 
$\hat{e}_{ij}= \frac{\vec{r}_i - \vec{r}_j}{|\vec{r}_i - \vec{r}_j|}$.
The first term in Eq.~(\ref{eq:dv_dt_grav}) is the gravitational force 
term usually used in SPH. The second term is due to gradients in the 
smoothing lengths and contains the quantities

\be
\zeta_k\equiv \frac{\partial h_k}{\partial \rho_k} \sum_j m_j \frac{\partial
  \phi_{kj}(h_k)}{\partial h_k}
\ee
and the $\Omega_k$ defined in Eq.~(\ref{eq:omega_i}). Formally, it looks very
similar to the pressure gradient terms in Eq.~(\ref{eq:momentum_equation_SPH})
with $G \zeta_k/2$ corresponding to $P_k/\rho_k^2$. As $\zeta_k$ is a negative
definite quantity, these adaptive softening terms act against the gas pressure
and therefore tend to increase the gravitational forces.\\
The explicit forms of $\phi$, $\phi'$ and $\partial \phi/\partial h$ for 
our cubic spline kernel can be found in Appendix A of \cite{price07a}. The
gravitational softening procedure obviously only applies to interacting SPH
particles. Generally, we use a binary tree based on \citet{benz90b} for the
long-range part of the gravitational forces. Depending on the choice of the
tree opening parameter, $\theta$, for each particle a list of nodes is
returned whose gravitational influences are calculated up to quadrupole
order. Forces from nearby, interacting particles are calculated via direct
summation according to the above prescription.

\subsubsection{Dissipation}
\label{sec:hydro_diss}

The conservation of mass, energy and momentum across a shock front requires
kinetic energy to be transformed into internal energy. Physically, this
transformation is mediated via viscosity and usually occurs over a length
scale of a few mean free paths in the gas. This length scale is generally
much shorter than any numerically resolvable length and thus in the numerical
discretization the transition appears to be a discontinuity. There are two
approaches to treat this problem, either by solving a local Riemann-problem 
as in Godunov-type methods, or by adding a controlled amount of viscosity
artificially to broaden the shock to a numerically resolvable width.
Not doing so results in unphysical oscillations in the post-shock region.
While the first approach is certainly more elegant, the second one 
is more robust and offers advantages in cases where the analytical solution 
to the Riemann problem is not known, for example, in the case of a 
complicated equation of state. Usually, the artificial viscosity approach 
is used in SPH and shock fronts are usually spread across a few smoothing 
lengths (rather than a few mean free paths) to make them numerically 
treatable.\\
We use an artificial viscosity prescription that is oriented at
Riemann-solvers \citep{monaghan97,price05} together with time dependent
viscosity parameters \citep{morris97,rosswog00} so that the dissipative terms
are only applied if they are really necessary to resolve a shock.
The additional term in the momentum equation reads

\be
\left(\frac{d\vec{v}_{i,\rm AV}}{dt}\right)= - \sum_j m_j 
\frac{\alpha^{\rm AV}_{ij}(t) v_{\rm sig} \vec{v}_{ij} \cdot 
\hat{e}_{ij}}{\rho_{ij}} \; \overline{\nabla_i W_{ij}},
\ee
where $\rho_{ij}= (\rho_i+\rho_j)/2$ and the symmetrized kernel gradient 
is given by

\be
\overline{\nabla_i W_{ij}}= \frac{1}{2}
\left[\frac{1}{\Omega_i} \nabla_i  W_{ij} (h_i) + 
\frac{1}{\Omega_j} \nabla_i  W_{ij} (h_j)\right]. 
\ee
The signal speed, $v_{\rm sig}$, is the fastest velocity with which information
can propagate between particle $i$ and $j$ and for the hydrodynamic case it is
given by

\be
v_{\rm sig}= \frac{c_i+c_j -  \vec{v}_{ij} \cdot \hat{e}_{ij}}{2},
\label{eq:vsig_hydro}
\ee
where $c_k$ is the sound velocity of particle $k$. The time dependent
parameter that controls the amount of artificial dissipation,
$\alpha^{\rm AV}_{ij}$, is 

\be
\alpha^{\rm AV}_{ij}= \frac{1}{4} \left\{ (\alpha_i(t)+\alpha_j(t)) 
\cdot (f_i+f_j) \right\},
\ee 
where the $f_k$ are the switches suggested by \cite{balsara95} to suppress
effects from artificial viscosity in pure shear flows

\be
f_k= \frac{|\nabla \cdot \vec{v}|_k}{|\nabla \cdot \vec{v}|_k+|\nabla \times
  \vec{v}|_k+ 10^{-4} \cdot c_k \rho_k/h_k}.\label{eq:balsara}
\ee
Here the small additive term in the denominator has been inserted to avoid
the switch from diverging in case both $|\nabla \cdot \vec{v}|$ and 
$|\nabla \times  \vec{v}|$ tend to zero.
The dissipative term in the evolution equation of the specific energy reads

\be
\left(\frac{du_{i,\rm AV}}{dt}\right)= - \sum_j m_j 
\frac{v_{\rm sig}}{\rho_{ij}} \left\{\frac{\alpha^{\rm AV}_{ij}}{2} 
\left[
\vec{v}_i \cdot \hat{e}_{ij} - \vec{v}_j \cdot \hat{e}_{ij}
\right]^2  \right\}
|\overline{\nabla_i W_{ij}}|.
\ee
It is straight forward to check that the total energy 
$\frac{d}{dt}(\sum_i \frac{1}{2} m_i \vec{v}_i^2 + m_i u_i)= 0$, 
i.e. that the applied dissipative terms are consistent with each 
other and conserve the total energy.\\
The viscosity coefficients, $\alpha_i$, are calculated according to
an additional evolution equation \citep{morris97}

\be
\frac{d\alpha_i}{dt}= -\frac{\alpha_i-\alpha_0}{\tau_i} + S_i,
\ee
where the decay constant is

\be
\tau_i = \frac{h_i}{0.2 \, c_i}
\ee
and the source term \citep{rosswog00}

\be
S_i= {\rm max}(-\nabla \cdot \vec{v}, 0) (2 - \alpha_i)
\ee
is used. In the absence of compression ($\nabla \cdot \vec{v} > 0$) the
parameter $\alpha_i$ decays to a minimum value $\alpha_0$, which we choose as
0.1. Note that this is more than an order of magnitude below the old
SPH-prescriptions that used values of a few, and it is further reduced by the
switch, Eq.~(\ref{eq:balsara}). In case of compression, $\nabla \cdot \vec{v}
< 0$, $\alpha$ can rise to values of up to 2 in the case of strong shocks. \\
Under certain circumstances it is desirable to add a small 
amount of thermal conductivity. This leads to an extra term in the evolution
equation of the specific energy

\be
\left(\frac{du_{i,\rm C}}{dt}\right)= 
- \sum_j m_j 
\frac{v_{\rm sig} \alpha^{\rm C}_{ij} (u_i-u_j)} {\rho_{ij}}
| \overline{\nabla_i W_{ij}} |,
\ee 
where $\alpha^{\rm C}_{ij}= (\alpha^{\rm C}_{i} + \alpha^{\rm C}_{j})/2$. 
The conductivity coefficient $\alpha^{\rm C}_{k}$ is evolved according to

\be
\frac{d\alpha^{\rm C}_k}{dt}= -\frac{\alpha^{\rm C}_k}{\tau_k} + S^{\rm C}_k,
\ee
where the decay constant is the same as above and the source term is given by
\citep{price05}

\be
S^{\rm C}_k = 0.1 h_k |\nabla^2 u_k|,
\ee
where we use the Brookshaw-type \citep{brookshaw85} second derivative

\be
\left(\nabla^2 u \right)_i= 2 \sum_j m_j \frac{u_i-u_j}{\rho_j} \frac{| \overline{\nabla_i W_{ij}} |}{r_{ij}}.
\ee

\subsection{Magnetic field}\label{sec:magnet}

The continuum equations to be solved are those of ideal magnetohydrodynamics.
The corresponding momentum equation reads
\be
\frac{dv^i}{dt}= \frac{1}{\rho} \frac{\partial S^{ij}}{\partial x^j}, 
\ee
where the stress tensor is given by
\be
S^{ij}= -P \delta^{ij} + \frac{1}{\mu_0} \left( B^i B^j - \frac{1}{2} B^2
  \delta^{ij} \right),
\ee
and the $B^k$ are the components of the magnetic field strength.
This form accounts for $\vec{B} (\nabla \cdot \vec{B})$ terms which 
are needed for momentum conservation in shocks but on the other hand are the 
cause of all the numerical instability, see \cite{price04a} for a detailed
discussion. Both the energy and the continuity equation have the same form as
in pure hydrodynamics, compare
Eqs.~(\ref{eq:energy_eq}) and (\ref{eq:continuity_eq}). \\
Here we present a discretized SPH-formulation including Euler potentials. 
This is our method of choice to evolve the magnetic field. 
For comparison and since some of the equations will be needed later, we also
summarize a more straightforward SPH discretization of the MHD equations due
to \citet{price05}. Both methods are implemented in the code and it is 
straightforward to switch between the two.

\subsubsection{Smoothed Particle Magnetohydrodynamics}\label{sec:SPMHD}
 
The following SPH discretization goes back to \citep{phillips85}, and has been 
extended and refined recently by \citet{price04a,price04b} and
\citet{price05}. This algorithm has been  extensively 
tested on a wide range of problems used to benchmark grid-based MHD codes. 
As in the hydrodynamic case, the formulation can be elegantly derived from 
a Lagrangian \citep{price04b}, guaranteeing the exact conservation of energy, 
entropy and momentum. The ``grad-$h$'' formulation of the SPMHD equations 
was derived in this manner by \citet{price04b} and we use this formulation 
here. The magnetic flux per unit mass $\vec{B}/\rho$ evolves according to

\begin{equation}
\frac{d}{dt}\left(\frac{\vec{B}_i}{\rho_{i}}\right) = -\frac{1}
{\rho_{i}^{2}\Omega_{i}} \sum_{j} m_{j} \vec{v}_{ij} \vec{B}_i\cdot 
\nabla_i W_{ij} (h_{i}),
\label{eq:ind}
\end{equation}
where $\Omega_i$ is the variable smoothing length term defined in 
Eq.~(\ref{eq:omega_i}).\\
The SPH formulation of the Lorentz force follows naturally from the 
Lagrangian and is given by \citep{price04b}

\begin{eqnarray}
\frac{d\vec{v}_{i, \rm mag,1}}{dt} = \hspace*{-0.7cm}&& -\sum_{j} \frac{m_{j}}
{\mu_{0}} 
\left\{ \frac{B_{i}^{2}/2}{\Omega_{i}\rho_{i}^{2}}\nabla_{i}W_{ij}(h_{i})
       +\frac{B_{j}^{2}/2}{\Omega_{j}\rho_{j}^{2}}\nabla_{i}W_{ij}(h_{j}) 
\right\}\nonumber \\
&& \hspace*{-1.8cm}+  \sum_{j} \frac{m_{j}}{\mu_{0}} \left\{ \frac{\vec{B}_{i} 
\left[\vec{B}_{i}\cdot \nabla_{i}W_{ij}(h_{i}) \right]} {\Omega_{i}
\rho_{i}^{2}} 
 + \frac{\vec{B}_{j} \left[\vec{B}_{j}\cdot \nabla_{i}
W_{ij}(h_{j}) \right]} {\Omega_{j}\rho_{j}^{2}}
\right\},  \label{eq:fcons}
\end{eqnarray}
where the terms correspond to the isotropic (due to gradients in 
magnetic pressure) and anisotropic magnetic force (due to field line 
tension) respectively. This exactly momentum-conserving form of the 
anisotropic SPMHD force is known to be unstable to a particle clumping 
instability in the regime where the magnetic field is dominant over the 
gas pressure (i.e. for tension forces) \citep{morris96a,morris96b,borve04}. 
Whilst typical magnetic field strengths found in compact objects mean 
that most simulations we will perform will lie in the regime where the 
above formulation is stable, a simple solution in the unstable regime is to
replace the anisotropic component of the magnetic force with a formulation
that vanishes for constant stress \citep{morris96a}, given by


\begin{eqnarray}
\frac{d\vec{v}_{i, \rm mag,2}}{dt} \hspace*{-0.1cm} = -\sum_{j} \frac{m_{j}}
{\mu_{0}} 
\left\{ \frac{B_{i}^{2}/2}{\Omega_{i}\rho_{i}^{2}}\nabla_{i}W_{ij}(h_{i})
       +\frac{B_{j}^{2}/2}{\Omega_{j}\rho_{j}^{2}}\nabla_{i}W_{ij}(h_{j}) 
\right\}\nonumber \\
 + \sum_{j} \frac{m_{j}}{\mu_{0}} \left\{ 
\frac{\vec{B}_{i} (\vec{B}_{i}\cdot \overline{\nabla_{i}W_{ij}}) 
- \vec{B}_{j} (\vec{B}_{j}\cdot \overline{\nabla_{i}W_{ij}})}
{\rho_{i} \rho_{j}} \right\}.
\label{eq:fmorr}
\end{eqnarray}
Using this force means that momentum is no longer conserved exactly
on the anisotropic term, however the effect of this small non-conservation
on shocks (where good conservation is critical) proves minimal (see
\cite{price04c}). 

\subsubsection{Dissipation}
\label{sec:mhd_diss}
 Dissipation terms necessary for the treatment of MHD shocks were 
formulated by \citet{price04a}. The induction equation contains a dissipative 
term corresponding to an artificial resistivity, ensuring that strong 
gradients in the magnetic field (i.e. current sheets) are resolved by 
the code. This term is given by

\begin{equation}
\frac{d}{dt}\left(\frac{\vec{B}_i}{\rho_{i}}\right) _{\rm dis,B} = \sum_{j} 
m_{j} \frac{\alpha^{\rm B}_{ij}(t) v_{sig}}{\rho_{ij}^{2}} (\vec{B}_{i} - 
\vec{B}_{j}) \vert \overline{\nabla W_{ij}} \vert,
\end{equation}
where the energy equation contains a corresponding term 

\begin{equation}
\left(\frac{du_i}{dt}\right) _{\rm dis,B} = \sum_{j} m_{j} 
\frac{\alpha^{\rm B}_{ij}(t) 
v_{sig}}{\mu_{0}\rho_{ij}^{2}} (\vec{B}_{i} - \vec{B}_{j})^{2} 
\vert \overline{\nabla W_{ij}} \vert. \label{eq:dudtavB}
\end{equation}
It is straightforward to demonstrate that this term gives a positive 
definite contribution to the entropy \citep{price04a}.

 In the magnetic field case we use a simple generalization of the 
signal velocity, Eq.~(\ref{eq:vsig_hydro}), given by
\begin{equation}
v_{sig} = \frac12({\rm v}_{i} + {\rm v}_{j}) - (\vec{v}_{ij}\cdot 
\vec{e}_{ij}),
\end{equation}
where ${\rm v}$ is the maximum propagation speed for MHD waves given by

\begin{equation}
{\rm v}_{i} = \frac{1}{\sqrt{2}} \left[ 
\left( c_{i}^{2} + v_{\rm A}^2\right)
 + \sqrt{ \left( c_{i}^{2}  + v_{\rm A}^2 
\right)^{2} - 4\frac{ c_{i}^{2}( \vec{B}_{i}\cdot\vec{e}_{ij} )^{2}}
{ \mu_{0}\rho_{i} }} 
\right]^{1/2},
\end{equation}
where $v_{\rm A}= \sqrt{\frac{B_{i}^{2}}{\mu_{0}\rho_{i}}}$ is the Alfv\'en
speed. 

\subsubsection{Euler potentials}\label{sec:euler_pots}
\begin{figure}
\psfig{file=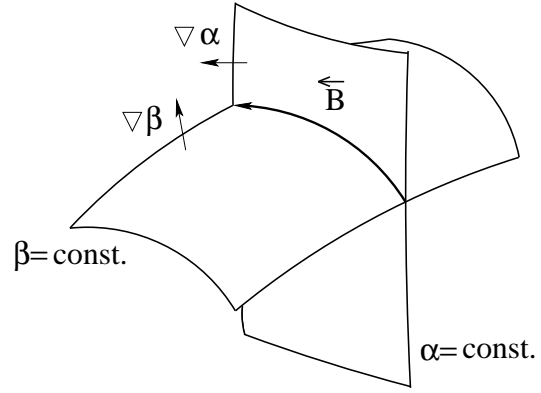,width=8cm}
\caption{Euler potentials: intersections of surfaces of constant values
  $\alpha$ and $\beta$ correspond to magnetic field lines.}
\label{fig:Euler potential}
\end{figure}

A key problem associated with the simulation of MHD phenomena is the 
maintenance of the divergence-free condition associated with the 
magnetic field. Whilst various methods for correcting the field 
produced by the standard SPMHD evolution Eq.~(\ref{eq:ind}) are possible 
\citep{price05}, we can avoid the problem entirely by 
formulating the magnetic field such that the divergence constraint is 
satisfied by construction. Use of the magnetic vector potential is one 
such construction. However for particle methods a natural choice is the
so-called `Euler potentials' (originally formulated by \citet{euler1769}-- 
see \citet{stern70}) but also referred to as the `Clebsch formulation' 
\citep[e.g.][]{phillips85}. In this formulation the magnetic field is
represented as

\begin{equation}
\vec{B} = \nabla\alpha \times \nabla\beta.
\label{eq:euler}
\end{equation}
Geometrically, the Euler potentials can be thought of as magnetic field line
labels \citep[e.g.][]{stern66}:  the magnetic field lines correspond to the
intersections of  surfaces of constant $\alpha$ with surfaces of constant
$\beta$, see Fig.~\ref{fig:Euler potential}. \\
The Euler potentials can be easily related to a vector potential which can be
of the form 
\be
\vec{A} = \alpha\nabla\beta + \nabla \xi
\ee
or
\be
\vec{A} = -\beta\nabla\alpha + \nabla \psi,
\ee
where $\xi$ and $\psi$ are arbitrary smooth functions.
It is straight forward to show that these vector potentials yield the B-field:
$\nabla \times \vec{A} = \nabla\alpha \times \nabla\beta = \vec{B}$. 
As the Euler potentials
only contain two independent variables (rather than the three components
of $\vec{A}$), they correspond to  an implicit choice of a gauge for the 
vector potential and that is maintained exactly during the further 
evolution. Taking the divergence of Eq.~(\ref{eq:euler}) demonstrates that 
the $\nabla\cdot\vec{B}=0$ constraint is satisfied by construction.\\
The condition that the magnetic field is frozen in translates into a
pure advection of the Euler potentials with the particles:
\begin{equation}
\frac{d\alpha_i}{dt} = 0, \hspace{1cm} \frac{d\beta_i}{dt} = 0.
\label{eq:advection}
\end{equation}
This advection property Eq.~(\ref{eq:advection}) means that the evolution of 
an arbitrary magnetic field can, in principle, be reconstructed from a 
hydrodynamic simulation given the initial and final particle positions 
(as long as the feedback from magnetic forces does not change the flow).\\ 
There are, however, some restrictions of the Euler potentials in comparison to
a MHD scheme where all three components of the magnetic field are evolved. 
These are: 

\begin{itemize}
\item[i)]  the calculation of the force involves second derivatives of the 
           potentials, which may be less accurate.
\item[ii)] zero\footnote{In our scheme there is some smoothing of the 
           Euler potential gradients in the calculation of $\vec{B}$ 
           from our use of Eqs.~(\ref{eq:grad_alpha}) and (\ref{eq:grad_beta})} 
           dissipation (i.e. no reconnection of field lines)
\item[iii)] non-linearity of initial conditions -- that is, for a given 
           $\vec{B}$ it is a non-trivial task to obtain the corresponding 
           Euler potentials.
\end{itemize}
 With regards to i) the tests presented here using the Euler potential 
formalism show no significant differences in force accuracy compared 
with similar tests shown using the standard SPMHD formalism. With regards 
to ii) we will demonstrate below how this restriction can be overcome by 
a simple modification to the Euler potential evolution. With regards to 
iii) the non-linearity of the Euler potentials may present some difficulty 
for the setting up of complicated initial conditions, but given the 
uncertainty of magnetic field configurations in most compact objects, 
we choose a simple initial configuration, so it does not present an 
immediate stumbling block for our simulations.\\
In two dimensions the Euler potentials are equivalent to a vector potential 
formulation with $\alpha = A_{z}$ and $\beta = z$.
We calculate the gradients of the Euler potentials in a way so that
gradients of linear functions are reproduced exactly, see e.g. 
\citet{price04c}. The gradient of the product of the density and 
an arbitrary quantity $A$ reads in standard SPH discretization
\be
[\nabla (\rho A)]_i= \sum_j m_j A_j \nabla_i W_{ij} (h_i) \label{eq:grad_rho_A}.
\ee
If we use the RHS of this equation on both to the left and the right side of
the equal sign and insert on the right the Taylor expansion of $A_j$ 
around $\vec{r}_i$
\be
A_j \approx A_i + (\vec{r}_j - \vec{r}_i)^\mu \left. \frac{\partial A}{\partial
  \vec{r}^\mu}\right\vert_{\vec{r}= \vec{r}_i},
\ee
one finds
\be
\sum_j m_j (A_j-A_i) \nabla_i^\nu W_{ij}=
\left. \frac{\partial A}{\partial \vec{r}^\mu} \right\vert_{\vec{r}=
\vec{r}_i} \sum_j m_j (\vec{r}_j - \vec{r}_i)^\mu \nabla_i^\nu W_{ij}, 
\ee
where we use Greek letters as summation indices, Latin ones for the
particle identities and the kernel gradients are evaluated with the smoothing
length $h_i$. This equation can be solved for the gradient of $A$ at the
position $\vec{r}_i$:


\be
\left. \frac{\partial A}{\partial
  \vec{r}^\mu}\right\vert_{\vec{r}= \vec{r}_i}= 
\left( \chi^{\mu \nu} \right)^{-1} \sum_j m_j (A_j-A_i) \nabla_i^\nu W_{ij}(h_i),
\ee
where the matrix $\chi^{\mu \nu}$ is given by
\be
\chi^{\mu \nu}= \sum_j m_j (\vec{r}_j - \vec{r}_i)^\mu \nabla_i^\nu
W_{ij}(h_i) 
\ee
and $\nabla_i^\mu$ is the $\mu$-component of the gradient evaluated at
position $\vec{r}_i$. Applied to the Euler potentials this yields
\begin{eqnarray}
(\nabla^\mu \alpha)_i= \left( \chi^{\mu \nu} \right)^{-1} \sum_j m_j (\alpha_j -
\alpha_i) \nabla_i^\nu W_{ij}(h_i)\label{eq:grad_alpha},\\
(\nabla^\mu \beta)_i= \left( \chi^{\mu \nu} \right)^{-1} \sum_j m_j (\beta_j -
\beta_i) \nabla_i^\nu W_{ij}(h_i)\label{eq:grad_beta}.
\end{eqnarray}
This formulation involves the inversion of a $3 \times 3$-matrix, $\chi$, for
each particle. This can be done analytically and the matrix only needs to be
stored 
for one particle at a time. The summations in Eqs.~(\ref{eq:grad_alpha}) and 
(\ref{eq:grad_beta}) do not involve densities, therefore they can be
conveniently be calculated in the density loop for subsequent use in the force
calculation. \\
Whilst in principle it is possible to formulate the magnetic forces 
using direct second derivatives of the Euler potentials -- e.g. making 
use of the SPH second derivative formulations of \citet{brookshaw85} 
generalized to vector derivatives by \citet{espanol03}-- it is not possible 
to do so and at the same time maintain the conservation of linear 
momentum in the force formulation, as the force involves a combination 
of first and second derivatives of the potentials which cannot be 
symmetrized. For this reason we simply use the usual force 
Eq.~(\ref{eq:fcons}) or Eq.~(\ref{eq:fmorr}) where $\vec{B}$ is the 
magnetic field computed using Eqs.~(\ref{eq:grad_alpha}), (\ref{eq:grad_beta})
and (\ref{eq:euler}).  
The tests presented in \S\ref{sec:mhd_tests} demonstrate that 
the resulting force is no less accurate than when the induction 
equation is used to evolve the magnetic field. A similar conclusion 
was reached by \citet{watkins96}, who, in formulating Navier-Stokes 
type viscosity terms for SPH, found that using a nested first 
derivative could in fact be more accurate than using \citet{brookshaw85} 
type terms.

\subsubsection{Euler potentials with dissipation}
\label{sec:eulerpots}
 The standard advection of the Euler potentials with the SPH particles 
results in zero dissipation of magnetic field lines and no reconnection. 
However, for problems involving shocks it is necessary to add dissipative 
terms that make the discontinuities numerically treatable by spreading
them over a few smoothing lengths. In order to do so we propose a simple 
modification of the Euler potential evolution based on the 
\citet{monaghan97} formulation of SPH dissipative terms

\begin{eqnarray}
\left(\frac{d\alpha_i}{dt}\right) _{\rm diss} & = & \sum_{j} m_{j} 
\frac{\alpha^B_{ij}(t) v_{sig}}{\rho_{ij}} (\alpha_{i} - \alpha_{j}) 
\vert \overline{\nabla W_{ij}} \vert,\label{eq:euler1diss} \\
\left(\frac{d\beta_i}{dt}\right) _{\rm diss} & = & \sum_{j} m_{j} 
\frac{\alpha^B_{ij}(t) v_{sig}}{\rho_{ij}} (\beta_{i} - \beta_{j}) 
\vert \overline{\nabla W_{ij}} \vert \label{eq:euler2diss},
\end{eqnarray}
which are SPH representations of the equations

\begin{equation}
\frac{d\alpha}{dt} = \eta \nabla^{2}\alpha; \hspace{1cm} 
\frac{d\beta}{dt} = \eta \nabla^{2}\beta.\label{eq:eulerdiss}
\end{equation}
where $\eta \sim \alpha^B_{ij} v_{sig} h$. Rigorous conservation of 
energy would require computing the resulting change in $\partial \vec{B}
/\partial t$ according to

\begin{equation}
\left(\frac{\partial \vec{B}}{\partial t}\right)_{diss} = \nabla 
\left(\frac{\partial \alpha}{\partial t}\right) \times \nabla  
\left(\frac{\partial \beta}{\partial t}\right), \label{eq:dBdteuler}
\end{equation}
which should be computed using the SPH formulations Eqs.~(\ref{eq:grad_alpha})
and  (\ref{eq:grad_beta})
for the gradients. However this would require an additional pass over 
the particles to compute the terms Eqs.~(\ref{eq:euler1diss}) and
(\ref{eq:euler2diss}) (after the density summation) before substituting 
the result into the SPH expression for Eq.~(\ref{eq:dBdteuler}). An 
approximate, but much more efficient solution is to simply compute the 
energy input according to Eq.~(\ref{eq:dudtavB}) using the $\vec{B}$ 
calculated from the Euler potentials. For the shock tube tests we find 
that this approximate approach is more than satisfactory.

Our sole purpose in formulating dissipative 
terms for the Euler potentials is to provide a mechanism to ensure that 
discontinuities in the magnetic field are treated appropriately by the 
numerical method (i.e. resolved over a few smoothing lengths). As such 
the terms given above do not, and are not intended to, correspond to a 
rigorous formulation of Ohmic dissipation using the Euler potentials. 
Indeed a more detailed derivation demonstrates that, whilst such a 
formulation would include terms of the form Eq.~(\ref{eq:eulerdiss}), 
additional terms would also be required in order for the dissipation 
to correspond meaningfully to the usual Ohmic dissipation terms added 
to the MHD induction equation. 

\subsection{Time integration}


The calculation of the gravitational forces is --together with the neighbor 
search-- the computationally most expensive task. It is 
therefore advantageous to choose a time integration method that only 
requires one force evaluation per time step.\\
The basic integration scheme that we use is a second order accurate
MacCormack predictor-corrector method \citep[e.g.][]{lomax01}. The predictor
is given by
\be
\tilde{y}_{i+1}= y_{i} + \Delta t_i y'_i,
\ee
the corrector is
\be
y_{i+1}= y_{i} + \frac{1}{2}\Delta t_i (y'_i + \tilde{y}'_{i+1}).
\ee
Here, $i$ labels the time step, the primes denote derivatives with 
respect to time, $\tilde{y}'_{i+1}$
denotes the derivatives at the predicted position and $\Delta t_i$ the 
used time step.\\
Our integration scheme allows for {\em individual time steps}, i.e.
each particle is evolved on its own timestep while the forces at
any given point in time, $t$, are calculated from the particle properties 
interpolated to $t$. At the beginning of the simulation, the ``desired'' time
step of each particle $i$ is determined
\be
\Delta t_{i, \rm des}= 0.2 \; {\rm min} \; (\Delta t_{f,i},\Delta t_{c,i}),
\label{eq:t_des}
\ee
where $\Delta t_{f,i}= \sqrt{h_i/|\vec{f}_i|}$ and $\vec{f}_i$ is the
particle's acceleration. The Courant-type time step is given by
$\Delta t_{c,i}= {\rm min}_j (h_j/v_{\rm sig,j})$ where $j$ runs over 
all neigbors and the particle $i$ itself. At the beginning of the 
simulation all particles start out with the same time step, $dt_0$, 
that is the maximum time step that fulfills the condition 
$dt_0< {\rm min}_i (\Delta t_{i, \rm des})$ and the condition that an integer
number of these time steps is equal to the time of the next data output. In
practice this means that the particles are running on time steps that are
slightly smaller than the desired ones, Eq.~(\ref{eq:t_des}). During the 
further evolution, we allow the particles to reduce their time step by 
a factor of $2^{-n}$, where $n$ is an integer, or, if the new desired 
time step is larger than twice the previously used time step, to increase 
the time step by a factor of 2. The
latter is only allowed if the next data output time can be hit exactly. 
Whenever a particle needs to be updated, all other particle properties are 
calculated at this point of time by interpolation to obtain the 
required derivatives.\\
The gain in computing time depends to a large extent on the application. In the
context of a neutron star merger the gain is only moderate, about a factor of
two. This is a consequence of the nearly incompressible neutron star matter
that results in a rather flat density distribution within the
stars. Therefore, the bulk of the particles has to be evolved on the shortest
time step. In other applications, however, say in the tidal
disruption of a star by a black hole \citep[e.g.][]{rosswog05a}, the gain can
be easily more than two orders of magnitude.

\section{Tests}
\label{sec:tests}
In this section we will describe tests of the new code elements. We start
with a description the initial particle setup and discuss why we use 
exclusively equal mass SPH particles for neutron star simulations. We 
then present tests of both the hydro- and the magnetohydrodynamics ingredients
in one two and three dimensions,
where the one and two dimensional tests are included as tests of the
algorithms for comparison with other codes and the three dimensional tests are
performed with the code itself.

\subsection{Particle setup}\label{sec:setup}

\begin{figure}
\psfig{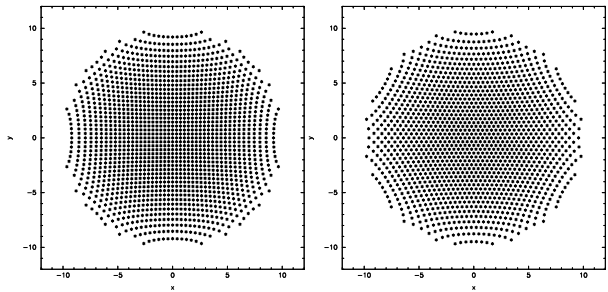}
\caption{Particle distribution in a slice of 0.5 code units thickness in 
the XY plane of stars with 50\,000 particles. The left panel shows 
the stretched cubic lattice, the right one the stretched 
close-packed configuration.}
\label{fig:cl_and_cp}
\end{figure}

\begin{figure}
\psfig{file=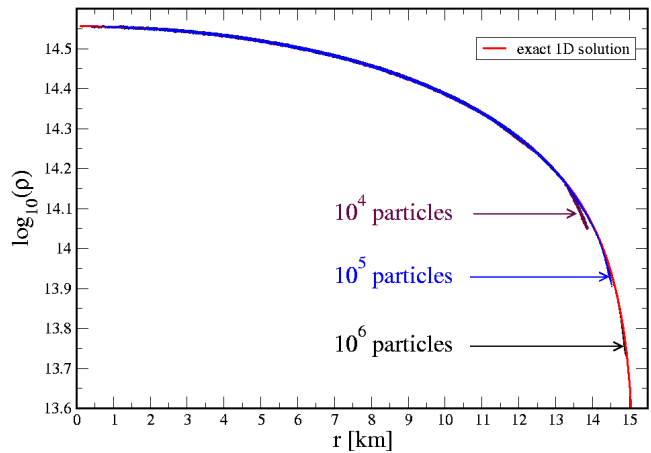,width=0.9\columnwidth,angle=-90}
\caption{Density profile of a 1.4 \msun $ $ neutron star. The solid red line
is the result of solving the 1D stellar structure equations, the dots are the
results obtained with the full SPH code at three different resolutions
($10^4$, $10^{5}$ and $10^6$ SPH particles).}
\label{fig:relaxation_on_profile}
\end{figure}

It is important to start a simulation from a SPH particle configuration 
that has been 'relaxed' into its optimal, minimal-energy configuration. 
This is usually
done with the full hydrodynamics code by applying an additional velocity 
dependent artificial acceleration term $\vec{f}_i \propto \vec{v}_i$. We 
discuss two different particle setups, a cubic lattice and a close-packed
configuration, in each case we only use particles of the same
mass. The case with unequal masses will be discussed below.\\
The first step is to solve the stellar structure equations in 1D to find 
the equilibrium profiles $\rho^{1D}(r), Y_e^{1D}(r)$ and $T^{1D}(r)$ for a 
neutron star of a specified mass, $M_{\rm ns}$. Here, $\rho$, $Y_e$ and $T$
are density, electron fraction and temperature. In the next step the 
desired number of particles, $N$, is distributed inside a unit sphere, 
either on a cubic lattice or as a close-packed configuration. To keep 
the particle mass constant, the number density of the particles, $n(r)$,  
has to reflect the density distribution of the star, $\rho^{1D}(r)= n(r) m$, 
where $m= M_{\rm ns}/N$  is the mass of each SPH particle. Subsequently
the unit sphere is stretched to the size of the neutron star so that the above
condition is met. The configuration constructed in this way is very close to
hydrostatic equilibrium.  Examples with both types of setups are shown in
Fig.~\ref{fig:cl_and_cp}.
To find the true numerical equilibrium state we relax this configuration 
with the full hydrodynamics code by applying an artificial, velocity 
dependent damping force \citep[e.g.][]{rosswog04b}. This procedure 
yields numerical equilibrium conditions with a minimal computational 
effort. To demonstrate that the particle distribution really settles 
to the correct result, we show in Fig.~\ref{fig:relaxation_on_profile} the
density profile of a 1.4 \msun $ $ neutron star as obtained by solving the stellar
structure equations in 1D (``exact'', red solid line). Overlaid are the
density distributions as obtained by relaxing three neutron stars of different
numerical resolution: maroon corresponds to $10\,000$, blue to $100\,000$ and 
black to $1\,000\,000$ SPH particles. The overall agreement with exact result 
is very good, deviations are only visible at the stellar edge where the 
extreme density gradients are challenging. \\
As a test of the quality of the initial conditions we set up the particles in
the described way, but instead of relaxing them, we evolve and
monitor the kinetic energy that builds up as a result of small deviations
from the true (numerical) hydrostatic equilibrium. We find only very minor
differences resulting from the different particle setups. For very low
particle numbers, say $1\,000$ particles, the gradients in the stellar profile
cannot be resolved properly and the particles adjust their positions to find
the equilibrium. 
For particle numbers in excess of $100\,000$ the particles smoothly move off the
initial grid but the overall density structure is practically unperturbed.\\
Different particle masses are known to introduce numerical noise into
SPH simulations. While a small range of particle masses may be admissible
in some applications, we only use equal particle masses. As a numerical
experiment, we set up a star with $10\,000$ particles and a constant  
particle number density, so that the particle masses carry the information 
of the stellar profile $\rho^{1D}(r)= n m(r)$. The extreme drop in density
towards the neutron star surface (caused by the very stiff equation of 
state) translates for this particle setup in particle masses that vary
by more than a factor of $10^6$. Without further relaxation we let this
configuration evolve. This worst case setup results in spurious particle 
motions as very light particles are in direct contact with heavier particles 
in the neutron star interior. The slightest noise of the heavy particles 
strongly disturbs the light ones (``ping pong on cannon ball effect'') and
leads to pathological particle densities, where low density particles can be
found in the interior of the star. Therefore, we only use equal mass
particles to keep the numerical noise at a minimal level.\\


\subsection{Hydrodynamics}\label{sec:hydro_tests}

\subsubsection{1D: Sod's shock tube}
\label{sec:sod}
\begin{figure}
\begin{center}
\psfig{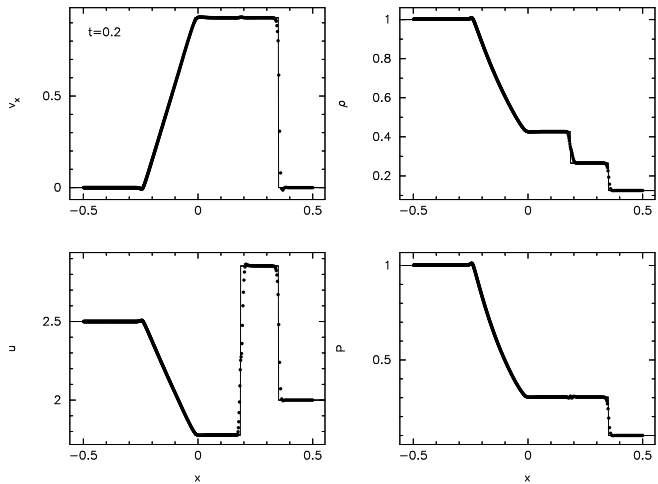}
\caption{Results of the Sod shock tube test in one dimension using 900 
SPH particles setup using unsmoothed initial conditions. Artificial 
viscosity and thermal conductivity are applied to appropriately smooth 
the shock and contact discontinuity respectively. The exact solution is 
given by the solid line. The upper row displays the
velocity (left) and the density (right), the bottom row shows specific
internal energy (left) and the pressure (right).}
\label{fig:sod1}
\end{center}
\end{figure}
As a standard test of the shock capturing capability we show 
the results of Sod's shock tube test \citep{sod78}. 
To the left of the origin, the initial state of the fluid is given by 
[$\rho, P, v_{x}$]$_{\rm L}$ = [1.0,1.0,0.0] whilst to the right of 
the origin the initial state is [$\rho, P, v_{x}$]$_{\rm R}$ = 
[0.125,0.1,0.0] with $\gamma = 1.4$. 
The problem is setup using 900 equal mass particles in one spatial dimension.
Rather than adopting the usual practice of smoothing the initial conditions 
across the discontinuity, we follow \citet{price04c} in using unsmoothed 
initial conditions but applying a small amount of artificial thermal 
conductivity using the switch described in \S\ref{sec:hydro_diss}. The results 
are shown in Fig.~\ref{fig:sod1}, where the points represent the SPH 
particles. For comparison the exact solution computed using a Riemann solver 
is given by the solid line.\\ 
The shock itself is smoothed by the artificial viscosity term, which in 
this case can be seen to spread the discontinuity over about 5 smoothing
lengths. 
The contact discontinuity is smoothed by the application of artificial 
thermal conductivity which (in particular) eliminates the ``wall heating'' 
effect often visible in numerical solutions to this problem. The exact 
distribution of particle separations in the contact discontinuity seen 
in Fig.~\ref{fig:sod1} is related to the initial particle 
placement across the discontinuity.\\
For this test, applying artificial viscosity and thermal conductivity as 
described, we do not find a large difference between the ``grad-$h$''
formulation and other variants of SPH based on averages of the smoothing 
length. If anything, the ``grad-$h$''-terms tend to increase the order of the 
method, which, as in any higher order scheme, tends to enhance oscillations 
which may otherwise be damped, visible in Fig.~\ref{fig:sod1} as 
small ``bumps'' at the head of the rarefaction wave (in the absence of 
artificial viscosity these bumps appear as small but regular oscillations 
with a wavelength of a few particle spacings).

\subsubsection{1D: The Einfeldt rarefaction test}
\label{sec:einfeldt}
\begin{figure}
\begin{center}
\psfig{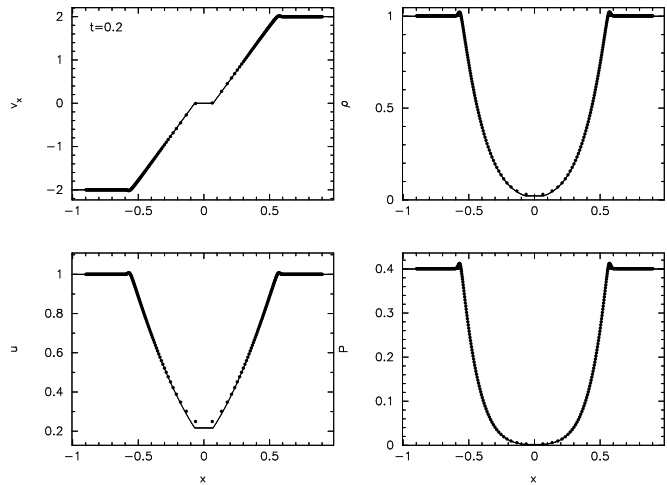}
\caption{Results of the Einfeldt test for the velocity (upper left), density
  (upper right), specific internal energy (lower left) and pressure (lower
  right). 400 particles were used in this test.}
\label{fig:Einfeldt}
\end{center}
\end{figure}
The initial conditions of the Einfeldt rarefaction test \citep{einfeldt91} 
do not exhibit any discontinuity in density or pressure, but the two halfs of
the computational domain move in opposite directions and thereby create a
region of very low density near the initial velocity discontinuity. This
low-density region 
represents a particular challenge for some iterative Riemann solvers which 
can return negative values for pressure/density. \citet{einfeldt91} designed
a series of tests  to illustrate this failure mode. The initial conditions of
this test are 
[$\rho, P, v_{x}$]$_{\rm L}$ = [1.0,0.4,-2.0] for the left state and [$\rho,
P, v_{x}$]$_{\rm R}$ = [1.0,0.4,2.0] for the right one. The results from a 400
particle calculation are shown in Fig.~\ref{fig:Einfeldt} after a time $t$=
0.2. The exact result is reproduced very accurately, only at the fronts of the
rarefaction waves a small overshoot occurs. The low density region does not
represent any problem for the method.

\subsubsection{3D: Sedov blast wave test}
\begin{figure*}
\begin{center}
\begin{turn}{270}\epsfig{file=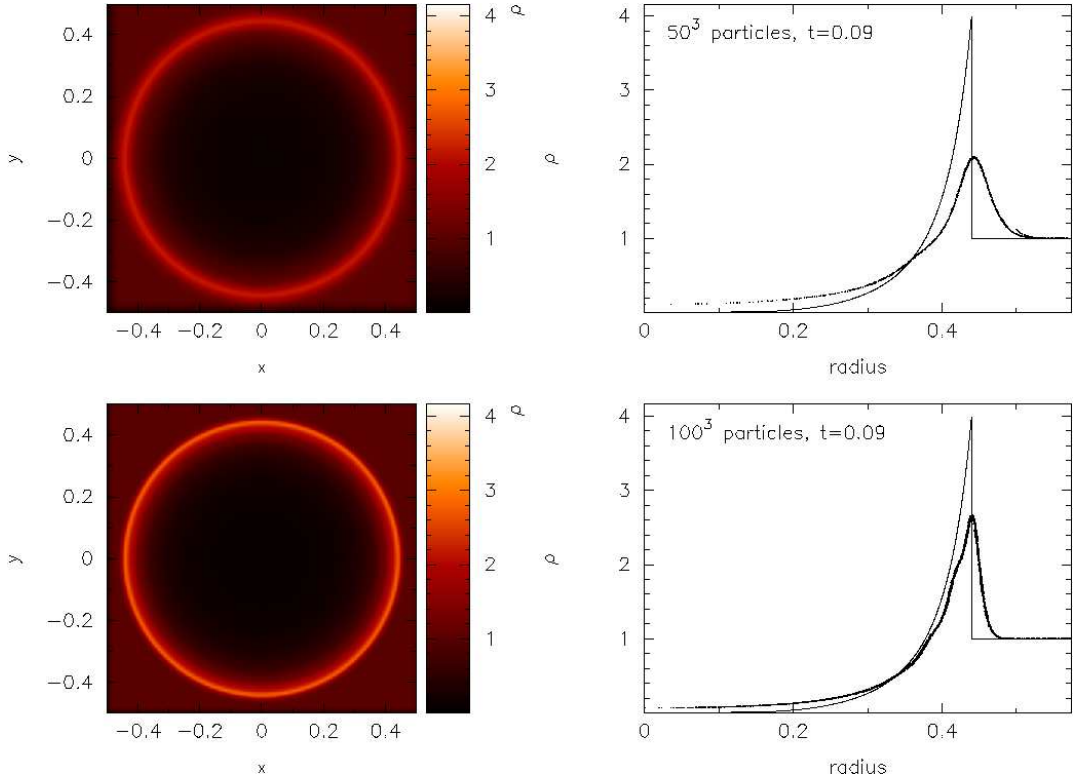, height=0.8\textwidth}\end{turn}
\caption{Results of the hydrodynamic Sedov blast wave test at 
$t=0.09$ at resolutions of $125\,000$ (top) and 1 million (bottom) 
particles respectively. The left panels show a rendered plot of the density in
a slice taken at $z=0$ whilst the right panels show the density and radial
position of each SPH particle, which may be compared to the exact solution
given by the solid line. Note that we are showing each individual SPH particle
and no averages.}
\label{fig:sedov}
\end{center}
\end{figure*}
In order to demonstrate that our scheme is capable of handling strong 
shocks in three dimensions, we have also tested the code on a Sedov 
blast wave problem both with, see Sec.~\ref{sec:MHD_blast}, and without
magnetic fields. Without  
magnetic fields the explosion is spherically symmetric, however for 
a strong magnetic field the blast wave is significantly inhibited 
perpendicular to the magnetic field lines, resulting in a compression 
along one spatial dimension. Similar tests for both hydrodynamics and 
MHD have been used by many authors -- for example by \citet{balsara01} 
in order to benchmark an Adaptive Mesh Refinement (AMR) code for MHD 
and by \citet{springel02} to test a new formulation (entropy equation)
of SPH.\\
The hydrodynamic version is set up as follows: The particles are 
placed in a cubic lattice configuration in a three dimensional domain 
$[-0.5,0.5] \times [-0.5,0.5] \times [-0.5,0.5]$ with uniform density $\rho =
1$ and zero  
pressure and temperature apart from a small region $r < R$ near the 
origin, where we initialize the pressure using the total blast wave 
energy $E=1$, ie. $P = (\gamma - 1) E/ (\frac43 \pi R^{3})$. We set 
the initial blast radius to the size of a single particle's smoothing 
sphere $R=2\eta \Delta x$ (where $2$ is the kernel radius, $\eta (= 1.5)$ 
is the smoothing length in units of the average particle spacing as in 
Eq.~(\ref{eq:h}) and $\Delta x$ is the initial particle spacing 
on the cubic lattice) such that the explosion is as close to point-like 
as resolution allows. Boundaries are not important for this problem, 
however we use periodic boundary conditions to ensure that the particle 
distribution remains smooth at the edges of the domain.\\
The results are shown in Fig.~\ref{fig:sedov} at $t=0.09$.
We have used a resolution of 50$^{3}$ and 100$^{3}$ particles 
(ie. $125\,000$ and 1 million particles respectively) and we have plotted
(left panels) the density in a $z=0$ cross section slice and (right panels)
the density and radial position of each particle (dots) together with the
exact self-similar Sedov solution (solid line).\\ 
We found that the key to an accurate simulation of this problem in 
SPH is to incorporate an artificial thermal conductivity term due to 
the huge initial discontinuity in thermal energy. The importance of 
such a term for shock problems in SPH has been discussed recently by 
\citet{price04c}. In the absence of this term the particle distribution 
quickly becomes disordered around the shock front and the radial 
profile appears to be noisy. From Fig.~\ref{fig:sedov} we see 
that at a resolution of 1 million particles the highest density in 
the shock at $t=0.09$ is $\rho_{\rm max}=2.67$ whereas for the lower 
resolution run $\rho_{\rm max} = 2.1$, consistent with a factor of 2 
change in smoothing length. Using this we can estimate that a 
resolution of $\sim345^{3} = 41$~million particles is required to 
fully resolve the density jump in this problem in three dimensions. 
Note that the minimum density obtained in the post-shock rarefaction 
also decreases with resolution. Some small-amplitude post-shock 
oscillations are visible in the solution which we attribute to 
interaction of the spherical blast wave with particles in the 
surrounding medium initially placed on a regular (Cartesian) cubic 
lattice. 


 
 
\subsubsection{3D: Radial oscillation of a neutron star}
\begin{figure}
\begin{center}
\begin{turn}{270}\epsfig{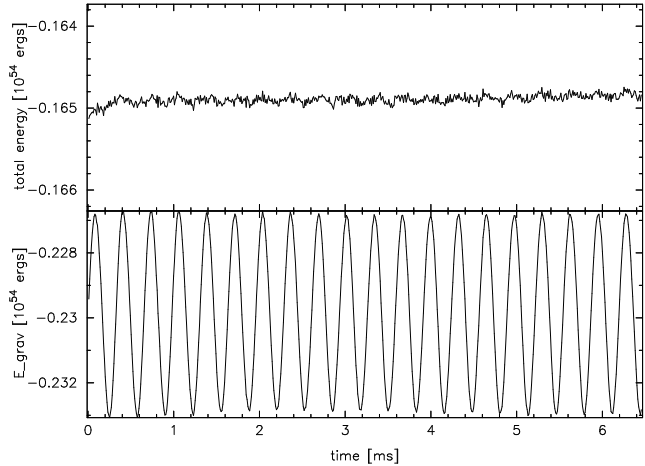}\end{turn}
\caption{Total (top) and gravitational potential energy (bottom) for 
10 radial oscillations of an initially hydrostatic neutron star using 
the Shen equation of state. }
\label{fig:nsoscills}
\end{center}
\end{figure}

As a further test case, we consider the radial oscillations of 
a neutron star using the Shen equation of state \citep{shen98a,shen98b}. The
initial conditions  
are a neutron star relaxed into hydrostatic equilibrium as described 
previously (\S\ref{sec:setup}), given an 
initial perturbation in velocity of the form ${\bf v} = v_{0} \hat{\bf r}$ 
where $v_{0}$ is an arbitrary but small amplitude (we choose $v_{0}= 0.01$c). 
No artificial viscosity or damping is applied for this problem since 
no shocks are involved. We compute the problem at low resolution 
using only $10\,000$ particles in the neutron star.\\
The results of this test are given in Fig.~\ref{fig:nsoscills}, which 
shows the results of an integration for 10 oscillation periods, where top 
and bottom panels show the total and gravitational potential energy 
respectively. From this Figure it may be observed that the amplitude is 
maintained almost exactly by the code over the 10 oscillation periods ($P
\approx 0.33$ ms) simulated. The residual 
fluctuations in total energy are directly attributable to a combination 
of the tree opening criterion (here $\theta=0.9$), which we find controls the 
level of ``noise'' in the total energy curve, and the timestepping 
accuracy (Courant number, here $C_{\rm cour}=0.2$), which affects the 
mean curve.\\
The fact that our SPH code, even at low resolution, is capable of 
following the neutron star oscillations for many periods without 
significant damping suggests that the code may be an ideal tool for 
studying neutron star oscillation modes. A similar study has recently 
been performed by \citet{monaghan06} who compared SPH simulations of the 
oscillation modes of two dimensional ``Toy stars'' \citep{monaghan04} with 
exact and perturbation solutions, finding good agreement between 
the two.

\subsubsection{3D: Binary orbit}

\begin{figure}
\psfig{file=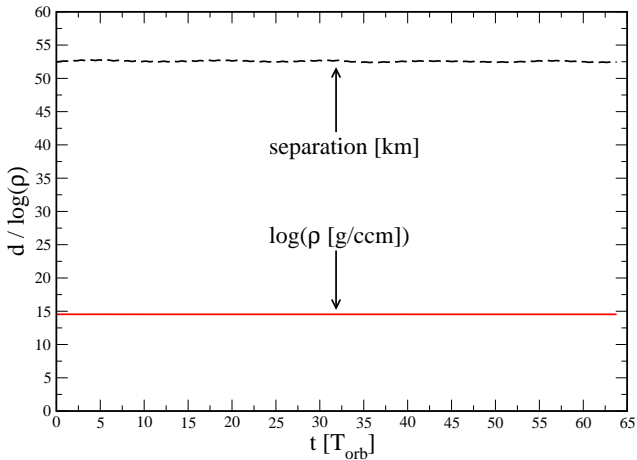,width=0.9\columnwidth,angle=-90}
\caption{Evolution of a neutron star binary system set on a circular
         orbit with a initial separation of $a_0=52.5$ km. Displayed
         are the orbital separation (upper line) and the maximum matter
         density (lower line). The binary system is evolved for as long as 
         63 orbital periods or about 920 neutron star dynamical time scales.}
\label{fig:orb_sep}
\end{figure}
\begin{figure}
\psfig{file=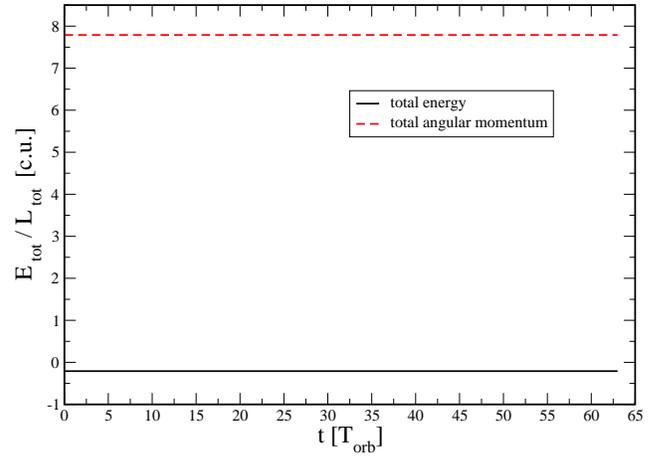,width=0.9\columnwidth,
angle=-90}
\caption{Evolution of total energy and angular momentum (both in code units) 
 during the orbital revolution of a binary system with twice 1.4  \msun.}
\label{fig:E_L_conservation}
\end{figure}
As a further test in 3D we set up a neutron star binary system on a stable
circular orbit and follow its long-term evolution. As initial separation
we choose $a_0= 52.5$ km (= 35 code units), no gravitational backreaction
forces are applied. To demonstrate that even at a very low resolution stable
and accurate orbital evolution can be obtained, we model each neutron star
with $1\,300$ SPH particles only. We relax two neutron stars in a corotating 
frame as described in \citet{rosswog04b}. After a tidally locked equilibrium 
configuration has been reached, the velocities are transformed to the 
space-fixed frame and the orbital evolution is followed with the full code
for as long as 63 orbital periods or approximately 920 dynamical time 
scales of the neutron stars. The evolution of the orbital separation of 
both neutron stars together with the maximum density in the binary system
are shown in Fig.~\ref{fig:orb_sep}. The binary stays nearly perfectly on 
the intended orbit. Due to the finite relaxation time very small scale
oscillations occur which lead to an exchange between orbital and oscillation
energy of the stars. This leads to small oscillations of the orbital 
separation around the initial value. But the corresponding deviations 
are very small ($\delta a_{\rm max}/a_0\approx 0.005$) and they do not grow 
during the very long evolution time. The central density is free of any
visible oscillation.\\
The evolution of the corresponding total energy, $E_{\rm tot}$, and the 
total angular momentum, $L_{\rm tot}$,
are shown in Fig.~\ref{fig:E_L_conservation}. Both quantities are excellently
conserved: $\delta E_{\rm tot}/E_{\rm tot,0}< 10^{-3}$ and 
$\delta L_{\rm tot}/L_{\rm tot,0}< 10^{-4}$.

\subsubsection{3D: Stellar head-on collision}
\begin{figure*}
\centerline{\psfig{file=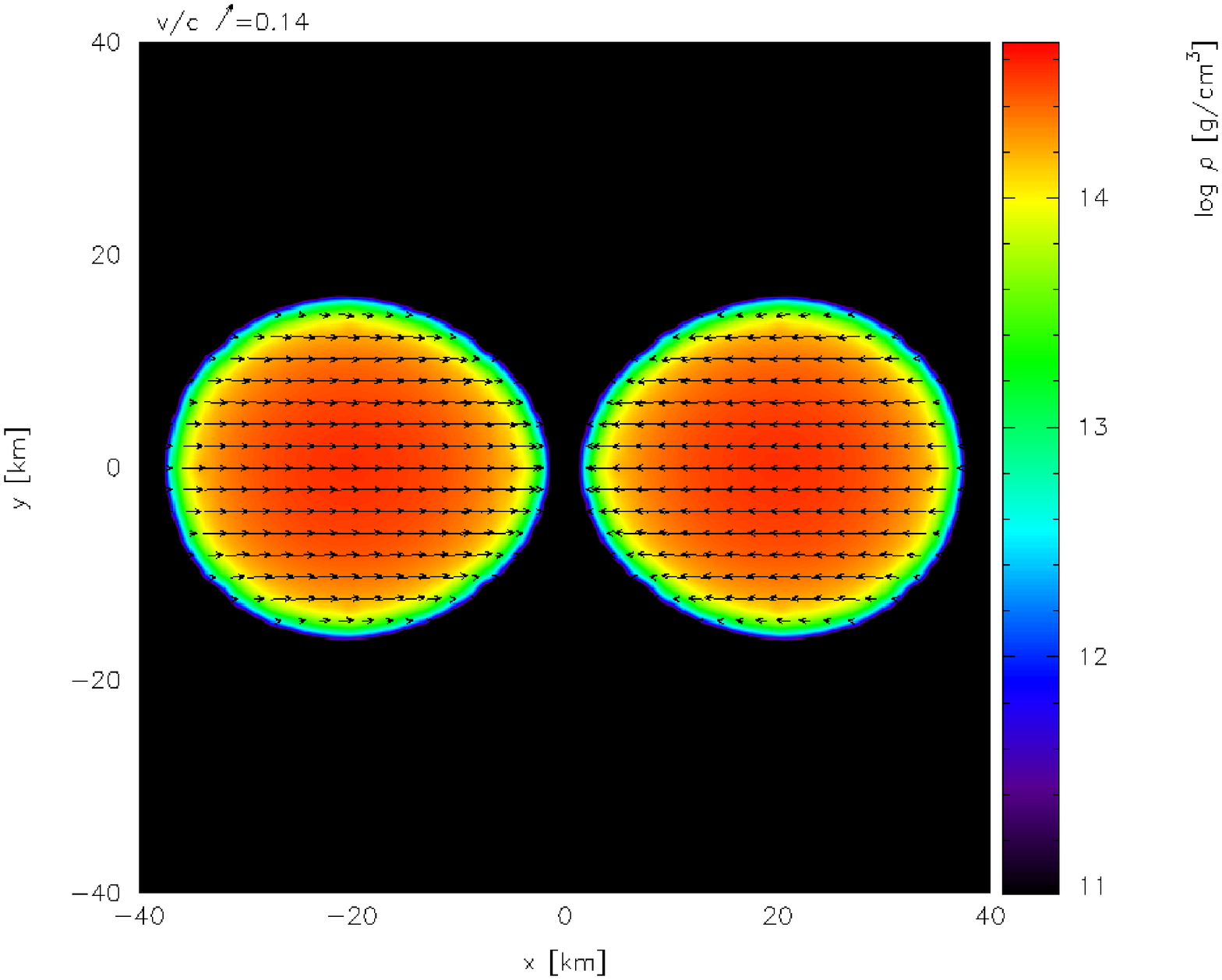,width=0.95\columnwidth,angle=-90}
\psfig{file=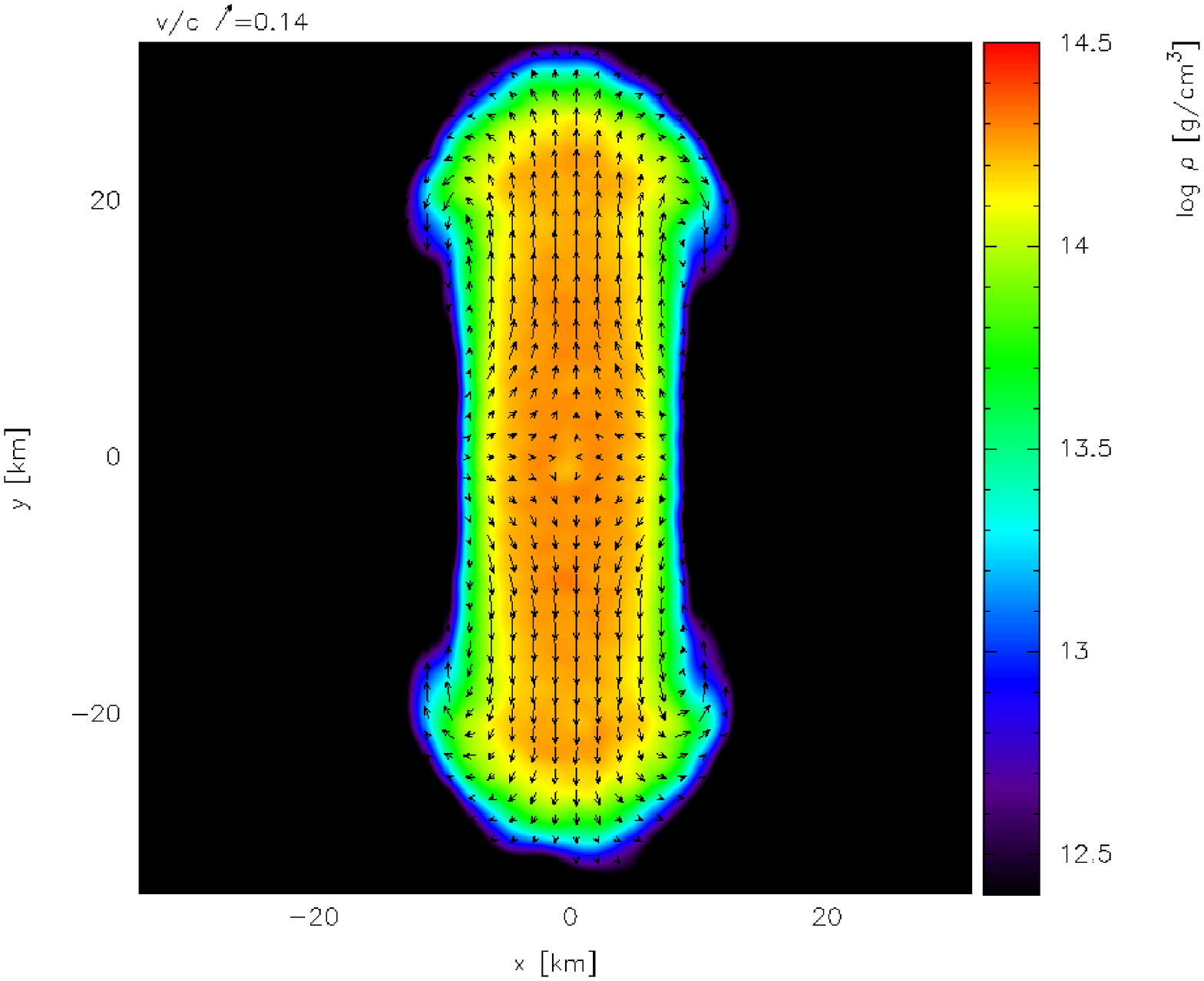,width=0.95\columnwidth,angle=-90}}
\caption{Head-on collision of two neutron stars. Such a head-on collision of
  two stars is considered a worst-case situation for the non-conservation of
  energy due to changes in the smoothing lengths, see text for details.}
\label{fig:head-on}
\end{figure*}
\begin{figure}
\psfig{file=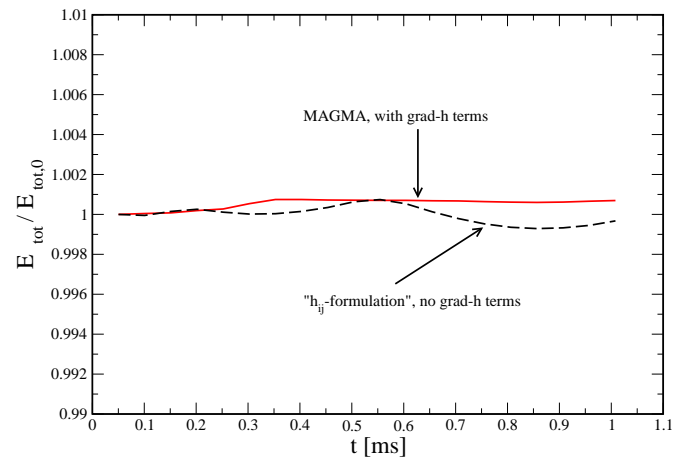,width=0.9\columnwidth,angle=-90}
\caption{Comparison of the energy conservation of both SPH-formulations (see
  text for details) for the above head-on collision. Both formulations yield
  very good conservation properties. The dominant error sources are the tree
  opening criterion and the time stepping criterion. To see a difference from
  the grad-h-terms, the tree opening criterion had to be reduced to
  $\theta=0.1$ and the pre-factor in Eq.~\ref{eq:t_des} to 0.1.}
\label{fig:energy_conservation_headon}
\end{figure}
It has long been known \citep{hernquist93} that using the SPH equations
derived under the assumption of constant smoothing lengths, e.g. in the
conventional $h_{ij}$-formulation summarized in Sec.~\ref{sec:hydro_gradh}, 
but still allowing the smoothing lengths to change in practice, can
in extreme cases lead to substantial errors in the conservation of energy.
For example, \citet{hernquist93} found a non-conservation of energy on a $\sim
10$-\%-level for a violent head-on collision of two polytropic stars.
To quantify this non-conservation for MAGMA we perform a similar head-on
collision between two neutron stars. Two neutron stars ($35\,000$ particles each)
of 1.4 \Mo obeying the \citet{shen98b} equation of state are set to an
initial separation of 35 code units (=52.5 km) and provided with an initial
relative velocity of 0.1 c. Fig.~\ref{fig:head-on} shows two density 
snapshots with overlaid velocity field during the evolution.\\
In Fig.\ref{fig:energy_conservation_headon} we show the evolution of the total
energy both for the $h_{ij}$-formulation and the new ``grad-$h$''-version 
(both are normalized to their initial values). As in previous
work \citep{rosswog02a,rosswog03a,rosswog03c} we use  on average 100 neighbors
for the $h_{ij}$-formulation, for the ``grad-$h$''-version we use a constant
of $\eta=1.5$ in Eq.~(\ref{eq:h}).\\
Generally the energy is very well conserved in both cases and the
non-conservation is determined by the tree-opening criterion and 
the time stepping accuracy. To see a difference between both formulations, we
reduced the tree opening criterion to $\theta=0.1$ and the pre-factor 
in Eq.~(\ref{eq:t_des}) to 0.1.
Both codes conserve energy in this challenging problem to better than
about $10^{-3}$  with the ``grad-$h$''-version showing a slightly better
performance. 
As in the other tests presented here, wee see a small improvement, but no
major change due to the use of the grad-$h$-terms.

\subsection{Magnetohydrodynamics}\label{sec:mhd_tests}
\subsubsection{1D: Brio-Wu shock tube test}
\begin{figure*}
\begin{center}
\begin{turn}{270}\epsfig{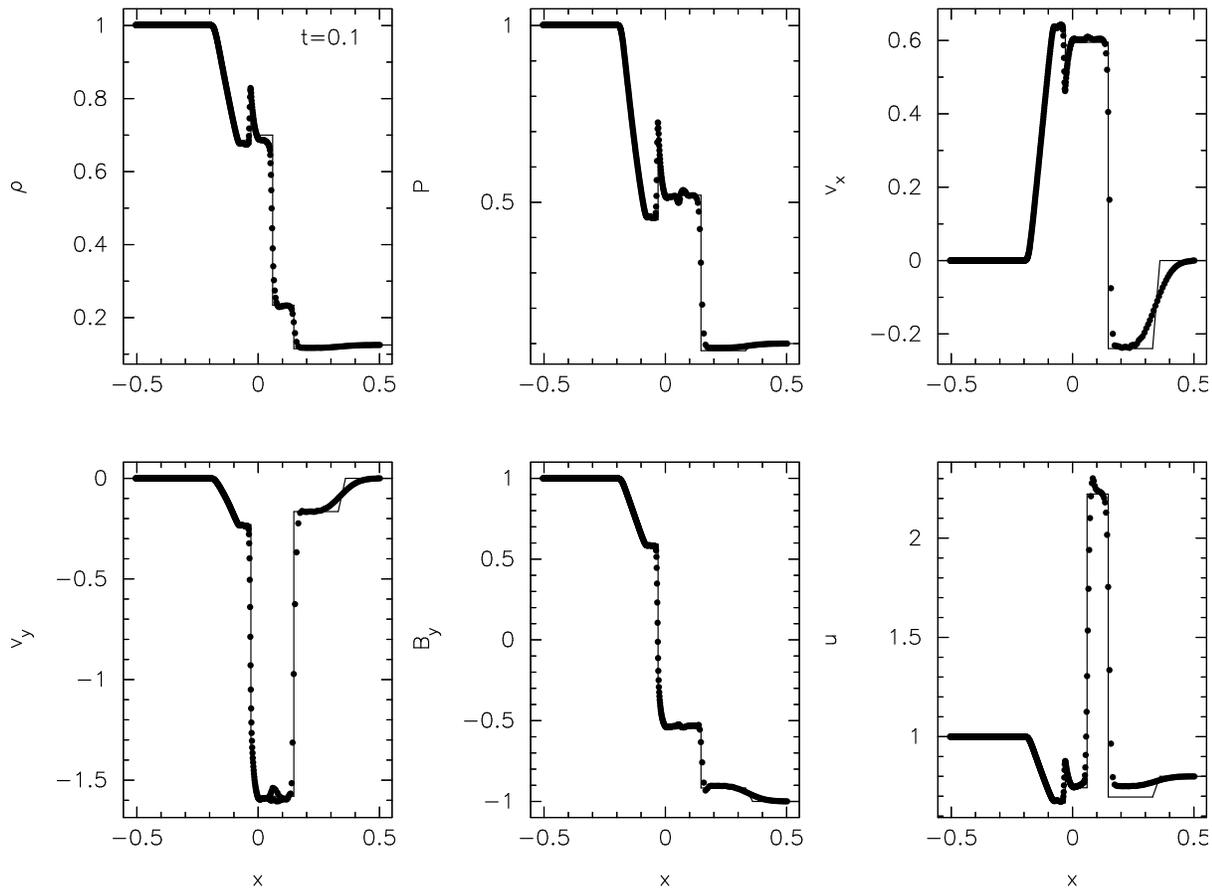}\end{turn}
\caption{Results of the Brio \& Wu MHD shock tube test at $t=0.1$ using 631 particles and the Euler potential formulation. For comparison the numerical solution taken from \citet{balsara98} is given by the solid line. The solution illustrates the complex shock structures which can be formed due to the different wave types in MHD, including in this case a compound wave consisting of a slow shock attached to a rarefaction wave. }
\label{fig:Brio_Wu}
\end{center}
\end{figure*}
The magnetic shock tube test of \cite{brio88} has become a standard test
case for numerical MHD schemes that has been widely used by many authors 
to benchmark (mainly grid-based) MHD codes 
\citep[e.g.][]{stone92a,dai94,ryu95,balsara98}.
The Brio-Wu shock test is the MHD analogon to 
Sod's shock tube problem that was described in Sec.~\ref{sec:sod}, 
but here no analytical solution is known. The MHD Riemann problem 
allows for much more complex solutions than the hydrodynamic case
which can occur because of the three different types of waves (i.e. 
slow, fast and Alfv\'en, compared to just the sound waves in hydrodynamics).
In the Brio-Wu shock test the solution contains
the following components (from left to right in Fig.~\ref{fig:Brio_Wu}): 
a fast rarefaction fan and a slow compound wave consisting of a slow 
rarefaction attached to a slow shock (moving to the left) and a contact 
discontinuity, a slow shock and a fast rarefaction fan (moving to the 
right). It has been pointed out, however, that the stability of the unusual
compound wave may be an artifact of the restriction of the symmetry to one 
spatial dimension whilst allowing the magnetic field to vary in two 
dimensions \citep{barmin96}.\\
Here we present the first results using the Euler potential formulation, see
\S\ref{sec:eulerpots}. Results of this problem using 
Smoothed Particle Magnetohydrodynamics (SPMHD) have been presented 
by \citet{price04a} and \citet{price04c}. The Euler potentials show a distinct 
improvement over the standard SPMHD results.
The initial conditions on the left side of the discontinuity are 
$[\rho,P,v_x,v_y,B_y]_{\rm L}= [1,1,0,0,1]$ and
$[\rho,P,v_x,v_y,B_y]_{\rm R}= [0.125,0.1,0,0,-1]$ on the right side. 
The $x-$component of the magnetic field is $B_{x}= 0.75$ everywhere and
a polytropic exponent of $\gamma = 2.0$ is used. Using the Euler potentials
the components are given  
by $\alpha = -B_{y} x$ (equivalent to the vector potential $A_{z}$) and 
$\beta = z$ (or more specifically $\nabla\beta = \hat{\bf z}$) and the 
$B_{x}$ component is treated as an external field which requires adding 
a source term to the evolution equation for $\alpha$ as discussed in 
\S\ref{sec:eulerpots}. Particles are restricted to move in one spatial 
dimension only, whilst the magnetic field is allowed to vary in two 
dimensions (that is, we compute a $v_{y}$ but do not use it to move the 
particles). This is sometimes referred to as a ``1.5D'' approximation.\\
We setup the problem using 631 equal mass particles in the domain 
$x \in [-0.5,0.5]$ using, as in the hydrodynamic case, purely discontinuous 
initial conditions. Artificial viscosity, thermal conductivity and 
resistivity are applied as described in \S\ref{sec:hydro_diss} and 
\S\ref{sec:eulerpots}. The results are shown at $t=0.1$ in 
Fig.~\ref{fig:Brio_Wu}. For comparison the numerical solution from 
\citet{balsara98} is given by the solid line (no exact solution exists 
for this problem). The solution is generally well captured by our 
numerical scheme. Two small defects are worth noting. The first is that a 
small offset is visible in the thermal energy -- this is a result of the 
small non-conservation introduced by use of the Morris formulation of the 
magnetic force (required for stability, see Eq.~(\ref{eq:fmorr})). Secondly,
the rightmost discontinuity  
is somewhat over-smoothed by the artificial resistivity term. We attribute 
this to the fact that the dissipative terms involve simply the maximum 
signal velocity $v_{sig}$ (that is the maximum of all the wave types). 
Ideally each discontinuity should be smoothed taking account of its 
individual characteristic and corresponding $v_{sig}$ (as would occur in 
a Godunov-MHD scheme). Increasing the total number of particles also 
decreases the smoothing applied to this wave.

\subsubsection{2D: Current loop advection problem}
A simple test problem for MHD is to compute the advection of a weak 
magnetic field loop. This test, introduced by \citet{gardiner05} in the 
development of the {\it Athena} MHD 
code\footnote{http://www.astro.princeton.edu/$\sim$jstone/athena.html}, 
presents a challenging problem for grid-based MHD schemes requiring 
careful formulation of the advection terms in the MHD equations. For 
our Lagrangian scheme, this test is straightforward to solve 
which strongly highlights the advantage of using a particle method 
for MHD in problems where there is significant motion with respect 
to a fixed reference frame.\\
We setup the problem here following \citet{gardiner05}: The computational 
domain is two dimensional with $x \in [-1, 1]$, $y \in [-0.5,0.5]$ 
using periodic boundary conditions. Density and pressure are uniform 
with $\rho=1$ and $P = 1$. The particles are laid down in a cubic 
lattice configuration with velocity initialized according to 
${\bf v} = (v_{0}\cos{\theta}, v_{0}\sin{\theta})$ with $\cos{\theta} 
= 2/\sqrt{5}$, $\sin{\theta} = 1/\sqrt{5}$ and $v_{0}=1$ such that by 
$t=1$ the field loop will have been advected around the computational 
domain once. The magnetic field is two dimensional, initialized using 
a vector potential given by
\begin{equation}
A_{z} = \alpha = \left\{ \begin{array}{ll}
A_{0}(R-r) & r \le R, \\
0 & r > R,
\end{array}\right.
\end{equation}
where $A_{0} = 10^{-3}$, $R=0.3$ and $r = \sqrt{x^{2} + y^{2}}$. The 
ratio of thermal to magnetic pressure is thus given by 
$\beta_{\rm plas} = P/(\frac12 B^{2}) = 2 \times 10^{6}$ (for $r < R$) such 
that the magnetic field is passively advected. 
\begin{figure}
\begin{center}
\begin{turn}{270}\epsfig{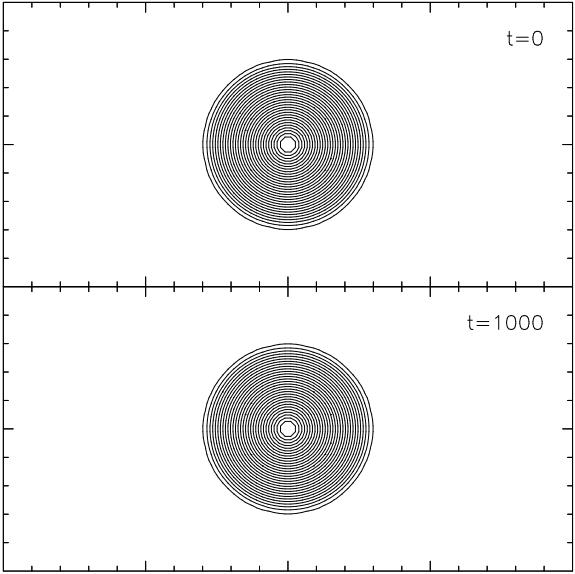}\end{turn}
\caption{Magnetic field lines in the current loop advection test, plotted at
  $t=0$ (top) and after $1\,000$ crossings of the computational domain (bottom).}
\label{fig:jadvect2D}
\end{center}
\end{figure}
\citet{gardiner05} show the results of this problem after two crossings of 
the computational domain, by which time the loop has either been 
significantly diffused or has disintegrated into oscillations depending 
on details of their particular choice of scheme. The advantage of a 
Lagrangian scheme is that advection is computed exactly, and using our 
Euler potential formulation (which in two dimensions is equivalent to a vector
potential formulation with $\alpha = A_{z}$ and $\beta = z$) for the magnetic
field, this is also true for the evolution  
of the magnetic field. The result is that the field loop is advected 
\emph{without change} by our code for as long as one may care to 
compute it. This is demonstrated in Fig.~{\ref{fig:jadvect2D} which 
shows the magnetic field lines at $t=0$ (top) and after $1\,000$ (this is not
a misprint!) 
crossings of the computational domain (bottom), in which the field 
configuration can be seen to be identical to the top figure. The magnetic 
energy (not shown) is also maintained exactly, whereas \citet{gardiner05} find 
of order a 10\% reduction in magnetic energy after two crossings of 
the domain.\\
In a realistic simulation involving MHD shocks there will be some 
diffusion of the magnetic field introduced by the addition of artificial 
diffusion terms, see Eq.~(\ref{eq:eulerdiss}), which are required to 
resolve discontinuities in the magnetic field. However the point is 
that these terms are explicitly added to the SPH calculation and can 
be turned off where they are not necessary (for example using the switches
described in \S\ref{sec:hydro_diss} and \S\ref{sec:mhd_diss}) whereas the
diffusion present in a grid-based code is intrinsic and always present.


\subsubsection{2D: Orszag-Tang test}

The evolution of the compressible Orszag-Tang vortex system \citep{orszag79} 
involves the interaction of several shock waves traveling at different 
speeds. Originally studied in the context of incompressible MHD turbulence,
it has later been extended to the compressible case
\citep{dahlburg89,picone91}. It is generally considered a good test 
to validate the robustness of numerical MHD schemes and has been used by many  
authors \citep[e.g.][]{ryu95,dai98,jiang99,londrillo00}. In the SPH context,
this test has been discussed in detail by \citet{price04c} and
\citet{price05}.\\ 
The problem is two dimensional with periodic boundary conditions on the 
domain $[0,1] \times [0,1]$. The setup consists of an initially 
uniform state perturbed by periodic vortices in the velocity field, which, 
combined with a doubly periodic field geometry, results in a complex 
interaction between the shocks and the magnetic field.\\
The velocity field is given by $\vec{v} = v_{0}[-\sin{(2\pi y)}, 
\sin{(2\pi x)}] $ where $v_{0} = 1$. The magnetic field is given by 
$\vec{B} = B_{0}[-\sin{(2\pi y)}, \sin{(4\pi x)}]$ where $B_{0} = 
1/\sqrt{4\pi}$. Using the Euler potentials this corresponds to 
$\alpha \equiv A_{z} = B_{0}/(2\pi) [ \cos{(2\pi y)} + 
\frac12\cos{(4\pi x)}]$. The flow has an initial average Mach number 
of unity, a ratio of magnetic to thermal pressure of $10/3$ and we 
use a polytropic exponent $\gamma = 5/3$. The initial gas state is therefore $P
= 5/3 B_{0}^{2}= 5/(12\pi)$ and $\rho = \gamma P/v_{0} = 25/(36\pi)$. Note
that the choice of length and time scales differs slightly between various 
implementations in the literature. The setup used above follows that 
of \citet{ryu95} and \citet{londrillo00}.\\
We compute the problem using $512\times 590$ particles initially placed on 
a uniform, close-packed lattice. The density at $t=0.5$ is shown in 
Fig.~\ref{fig:orszagtang} using both the standard SPMHD formalism 
(left), see \S\ref{sec:SPMHD}, and the Euler potential formalism (right), see
\S\ref{sec:euler_pots}. The Euler potential formulation is clearly superior
to the standard SPMHD method. This is largely a result of the relative 
requirements 
for artificial resistivity in each case. In the standard SPMHD method the 
application of artificial resistivity is crucial for this problem (that is, 
in the absence of artificial resistivity the density and magnetic field 
distributions are significantly in error). Using the Euler potentials we 
find that the solution can be computed using zero artificial resistivity, 
relying only on the ``implicit smoothing'' present in the computation of 
the magnetic field using SPH operators in Eqs.~(\ref{eq:grad_alpha}) and
(\ref{eq:grad_beta}). This means  
that topological features in the magnetic field are much better preserved, 
which is reflected in the density distribution. For example the filament 
near the center of the figure is well resolved using the Euler potentials 
but completely washed out by the artificial resistivity in the standard 
SPMHD formalism. Also the high density features near the top and bottom 
of the figure (coincident to a reversal in the magnetic field) are much 
better resolved using the Euler potentials.\\
A further advantage of using the Euler potentials is that the field lines 
can be plotted directly as equipotential surfaces of the potentials. The 
field lines corresponding to Fig.~\ref{fig:orszagtang} are thus shown 
in Fig.~\ref{fig:fieldlines}.\\
In order to enable a comparison between different codes, we also show
the 1D pressure distribution along the lines 
$y=0.4277$ and $y=0.3125$ in Fig.~\ref{fig:orszagtangcuts} which may 
be compared to similar plots given in Fig.~11 of \citet{londrillo00}
and in \citet{jiang99}.
\begin{figure*}
\begin{flushleft}
\begin{turn}{270}\epsfig{file=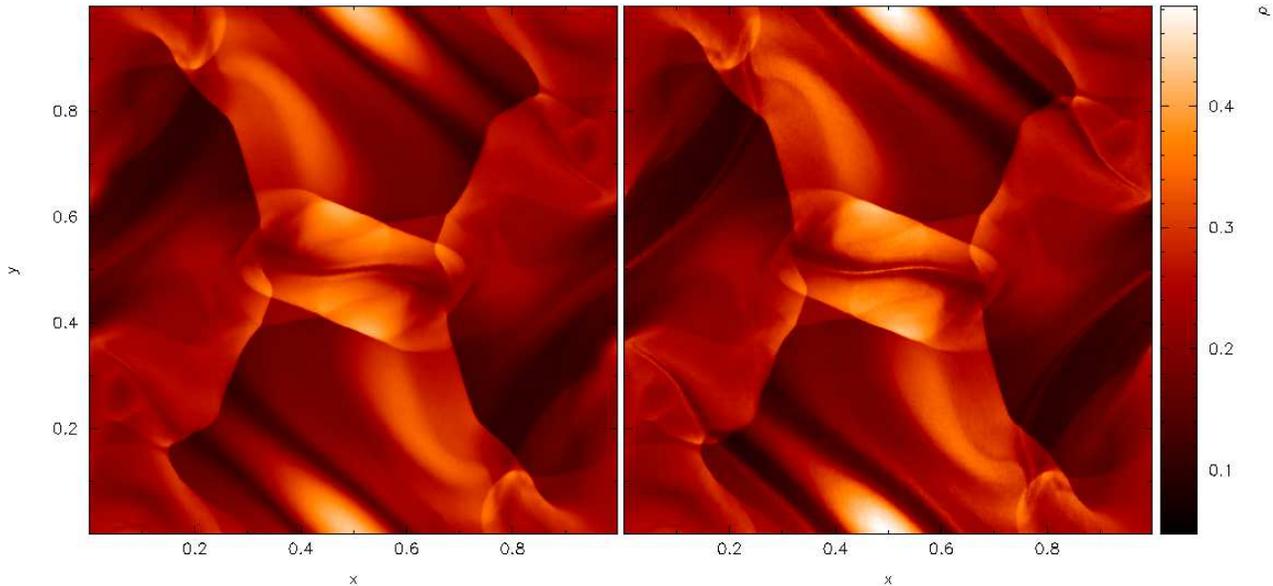,height=0.95\textwidth}\end{turn}
\caption{Density distribution in the two dimensional Orzsag-Tang vortex 
problem at $t=0.5$. The initial vortices in the velocity field combined 
with a doubly periodic field geometry lead to a complex interaction 
between propagating shocks and the magnetic field. Results are shown 
using $512\times 590$ particles using a standard SPMHD formalism (left) 
and using the Euler potentials (right). The reduced artificial 
resistivity required in the Euler potential formalism leads to a much 
improved effective resolution. The plot may be compared to results 
shown at comparable resolution in Fig.~14 of Dai \& Woodward (1998).}
\label{fig:orszagtang}
\end{flushleft}
\end{figure*}

\begin{figure}
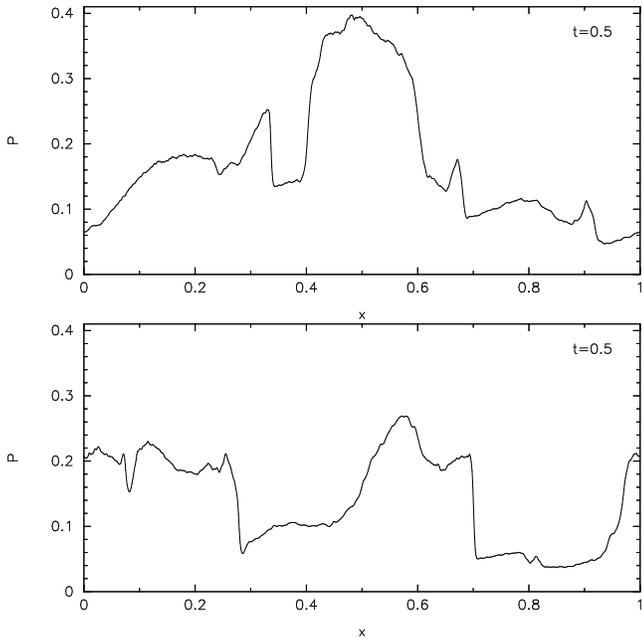

\begin{center}
\begin{turn}{270}\epsfig{file=orszagtang_cut_y04277.ps, 
height=\columnwidth}\end{turn}
\begin{turn}{270}\epsfig{file=orszagtang_cut_y03125.ps, 
height=\columnwidth}\end{turn}
\caption{Pressure distribution in the Orzsag-Tang vortex problem (as in 
Fig.~\ref{fig:orszagtang}) along a 1D cut taken at $y=0.4277$ (upper 
panel) and $y=0.3125$ (lower panel). The plots can be compared to 
Fig.~11 in \citet{londrillo00}, see also \citet{jiang99}.}
\label{fig:orszagtangcuts}
\end{center}
\end{figure}

\begin{figure}
\begin{flushleft}
\begin{turn}{270}\epsfig{file=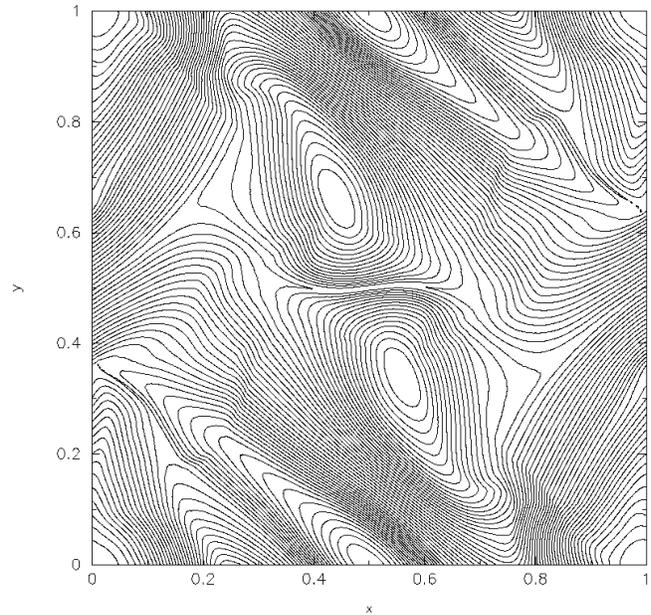, 
height=\columnwidth}\end{turn}
\caption{Magnetic field lines (contours of the Euler potential 
$\alpha$) in the two dimensional Orszag-Tang vortex problem at $t=0.5$ 
(corresponding to the right panel of Fig.~\ref{fig:orszagtang}). This 
may be compared to Fig.~15 in \citet{dai98} and Fig.~10 in \citet{londrillo00}.}
\label{fig:fieldlines}
\end{flushleft}
\end{figure}

\subsubsection{3D: MHD blast wave}\label{sec:MHD_blast}
\begin{figure*}
\begin{center}
\begin{turn}{270}\epsfig{file=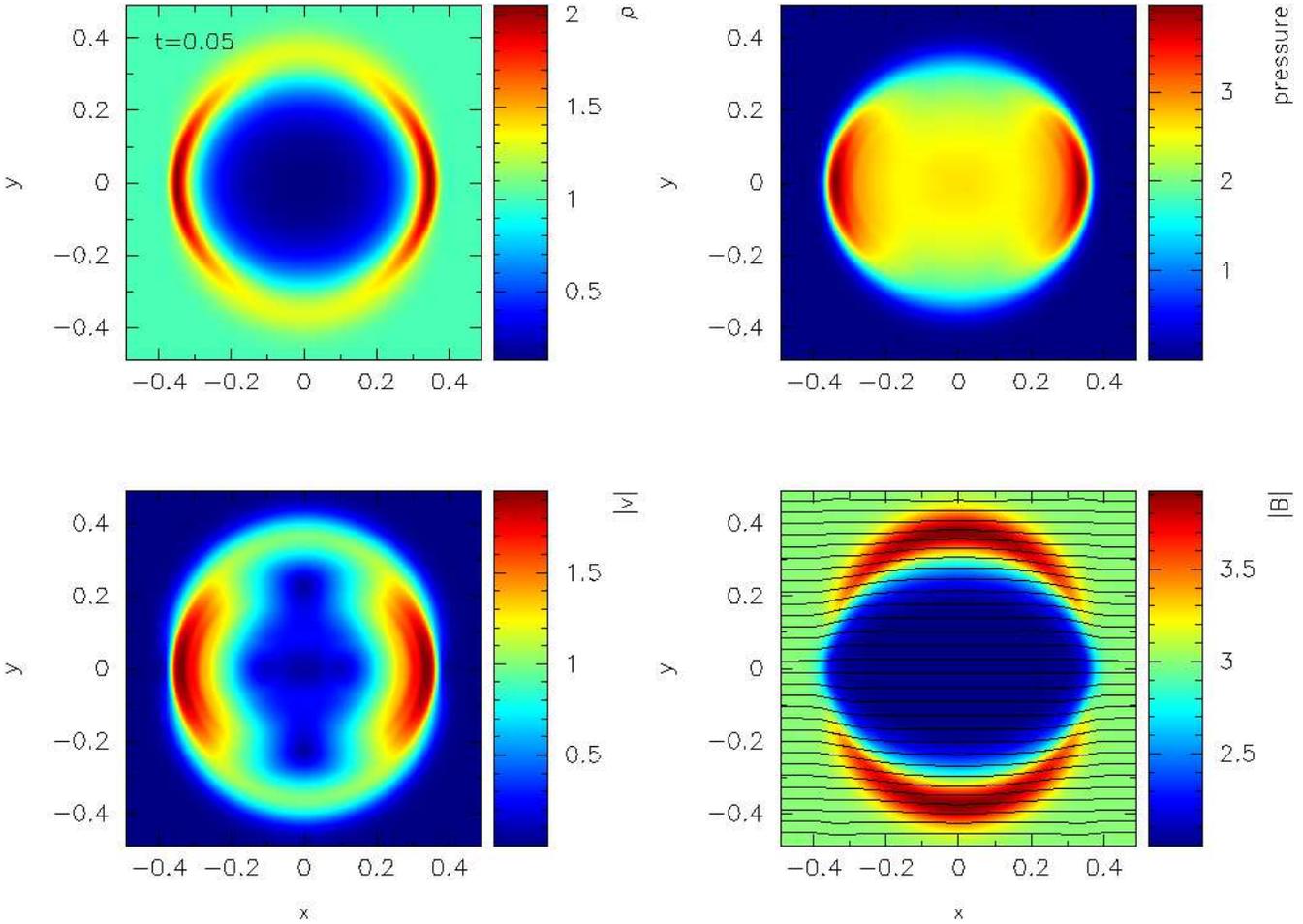, height=\textwidth}\end{turn}
\caption{Results of the MHD blast wave test at $t=0.05$ at a 
resolution of 1 million ($100^{3}$) particles. Plots show (left 
to right, top to bottom) density, pressure, magnitude of velocity 
and magnetic field strength (with overlaid field lines), plotted in 
a cross-section slice through the $z=0$ plane.}
\label{fig:sedovmhd}
\end{center}
\end{figure*}
There appear to be very few three dimensional MHD solutions published  
in the literature. Here we perform an MHD version of the Sedov test,  
identical to the hydrodynamic test with the addition of a uniform magnetic
field in the $x-$ direction, that is ${\bf B} = [B_{0},0,0]$ with $B_{0} =
3.0$. A similar test has been used by \citet{balsara01} for testing a 3D  
Adaptive Mesh Refinement code although with weak magnetic fields. Here  
we perform the test in the strong field regime such that the geometry  
of the blast is significantly constrained by the magnetic field,  
testing both the magnetic field evolution and the formulation of  
magnetic forces in the code. Initially the  
surrounding material has zero thermal pressure, meaning that the 
plasma $\beta_{\rm plas}$ is zero (ie. magnetic pressure infinitely strong 
compared to thermal pressure). However, this choice of field strength 
gives a mean plasma $\beta_{\rm plas}$ in the post-shock material of 
$\beta_{\rm plas}\sim 1.3$, such that the magnetic pressure plays an equal or
dominant role in the evolution of the shock.\\  
The initial Euler potentials for the blast wave are:
\be
\alpha = -B_0 z \quad {\rm and} \quad 
\beta = y
\ee
where an offset is applied to each potential at the boundaries to  
ensure periodicity.\\
The results of this problem at $t=0.05$ are 
shown in Fig.~\ref{fig:sedovmhd}, where plots show density, pressure, 
magnitude of velocity and magnetic field strength  in a 
cross section slice taken at $z=0$. In addition the magnetic field 
lines are plotted on the magnetic field strength plot. \\
In this strong-field regime, the magnetic field lines are not 
significantly bent by the propagating blast wave but rather strongly 
constrain the blast wave into an oblate spheroidal shape. The density 
(and likewise pressure) enhancement in the shock is significantly 
reduced in the $y-$direction (left and top right panels) due to the 
additional pressure provided by the magnetic field which is compressed 
in this direction (bottom right panel).

\section{Summary and Outlook}
\label{sec:summary}
We have introduced a new, 3D code, MAGMA, for astrophysical
magnetohydrodynamic problems that is based on the smoothed particle
hydrodynamics method. 
The equations of self-gravitating hydrodynamics  are derived
self-consistently from a Lagrangian and account in particular for the so-called
``grad-h''-terms. Contrary to other approaches, we also account for the extra
terms in the gravitational acceleration terms that stem from changes in the
smoothing length. This part of the code has been extensively tested on a
large set of standard test problems. The code performs very well, in particular
its conservation properties are excellent. While the ``grad-h''-terms slightly
improve the accuracy, in typical applications involving neutron stars the
differences to the older set of equations are very minor.\\ 
We evolve the magnetic fields with so-called Euler potentials which are
advected on the SPH-particles. They correspond to a formulation of the
magnetic field in terms of a vector potential, therefore, the
$\nabla\cdot\vec{B}=0$ constraint is satisfied by construction. 
To handle strong shocks artificial dissipative terms were introduced in 
these potentials, but for several tests no artificial dissipation is
required and the corresponding terms can be switched off. The Euler potential
approach shows in all tests a considerably higher accuracy than previous
magnetic SPH formulations and is our method of choice for our future
astrophysical applications of the MAGMA code.

\section*{Acknowledgements}

It is a pleasure to thank Joachim Vogt for many enlightening discussions and
for bringing the Euler potentials to our attention.\\
DJP is supported by a UK PPARC postdoctoral research
fellowship. Visualizations and exact solutions were computed using SPLASH, an
interactive visualization tool for SPH publicly available from
http://www.astro.ex.ac.uk/people/dprice/splash.\\ 
Part of the simulations presented in this paper were performed on the JUMP
computer of the H\"ochstleistungsrechenzentrum J\"ulich.


\begin{appendix}

\end{appendix}

\bibliography{astro_SKR}   
\bibliographystyle{mn2e.bst}

\end{document}